\documentclass[a4paper,fleqn,usenatbib]{mnras}

\usepackage{newtxtext,newtxmath}

\usepackage[T1]{fontenc}
\usepackage{ae,aecompl}

\usepackage{graphicx}    
\usepackage{amsmath}    
\usepackage{amssymb}    
\usepackage{subcaption}
\usepackage{bm}

\newcommand{\kms}{km\,s$^{-1}$}
\newcommand{\mpch}{\,$h^{-1}$\,Mpc}
\newcommand{\hmpc}{\,$h$\,Mpc$^{-1}$}
\newcommand{\gpch}{$h^{-1}$\,Gpc}
\newcommand{\msol}{$h^{-1}$\,M$_{\sun}$}
\newcommand{\hubbleunit}{$h$\,km\,s$^{-1}$\,Mpc$^{-1}$}

\title[Density-velocity cross-correlation in 6dFGS]{Improving constraints on the growth rate of structure by modelling the density-velocity cross-correlation in the 6dF Galaxy Survey}

\author[Adams and Blake]{
	Caitlin Adams$^{1,2}$\thanks{E-mail: cadams@swin.edu.au}
	and Chris Blake$^{1,2}$
	\\
	$^{1}$Centre for Astrophysics \& Supercomputing, Swinburne University of Technology, P.O. Box 218, Hawthorn, VIC 3122, Australia.\\
	$^{2}$ARC Centre of Excellence for All-sky Astrophysics (CAASTRO)
}

\date{Accepted 2017 June 15. Received 2017 June 13; in original form 2017 March 28}

\pubyear{2017}

\begin{document}
	\label{firstpage}
	\pagerange{\pageref{firstpage}--\pageref{lastpage}}
	\maketitle
	
	\begin{abstract}
		We present the first simultaneous analysis of the galaxy overdensity and peculiar velocity fields by modelling their cross-covariance. We apply our new maximum-likelihood approach to data from the 6-degree Field Galaxy Survey (6dFGS), which has the largest single collection of peculiar velocities to date. We present a full derivation of the analytic expression for the cross-covariance between the galaxy overdensity and peculiar velocity fields and find direct evidence for a non-zero correlation between the fields on scales up to $\sim$50\mpch. When utilising the cross-covariance, our measurement of the normalised growth rate of structure is $f\sigma_8(z=0) = 0.424^{+0.067}_{-0.064}$ (15\% precision), and our measurement of the redshift-space distortion parameter is $\beta=0.341^{+0.062}_{-0.058}$ (18\% precision). Both measurements improve by $\sim$20\% compared to only using the auto-covariance information. This is consistent with the literature on multiple-tracer approaches, as well as Fisher matrix forecasts and previous analyses of 6dFGS. Our measurement of $f\sigma_8$ is consistent with the standard cosmological model, and we discuss how our approach can be extended to test alternative models of gravity.
	\end{abstract}
	
	
	
	
	\section{Introduction}\label{sec:introduction}
	Our understanding of the Universe can be neatly summarised by the standard cosmological model: under the assumption of a Friedmann-Lema\^{i}tre-Robertson-Walker metric governed by general relativity, the Universe is flat and its total energy is distributed between baryonic matter ($\sim$5\%), dark matter ($\sim$25\%), and dark energy in the form of the cosmological constant ($\sim$70\%). This picture has been refined over many years through extensive testing against different observables, including the cosmic microwave background \citep{PlanckCollaboration2015}, type Ia supernovae \citep{Betoule2014}, and the distribution of large-scale structure \citep{Anderson2014}. Each observable tests different features of the model, and together they have established a robust framework for cosmology. This model is commonly labelled as $\Lambda$ cold dark matter ($\Lambda$CDM), which is shorthand for two of its assumptions: the majority of energy in our universe is either CDM or an energy density in the form of the cosmological constant ($\Lambda$). While $\Lambda$ was originally inserted into Einstein's equations to keep the Universe from collapsing, it has come to represent the mechanism behind accelerating expansion, which was first observed by \cite{Riess1998} and \cite{Perlmutter1999}. Despite the fact that its contribution to the total energy density is very well constrained by modern studies, its physical form is poorly understood. 
	
	There are two prominent lines of thinking regarding the cause of accelerating expansion: it is either driven by an energy density with negative pressure (specifically the cosmological constant, or more generally dark energy), or it is a natural consequence of gravity deviating from general relativity on large scales, as described by modified gravity models. See \cite{Joyce2016} for an excellent review of both.
	
	Any modification to gravity will affect the formation of large-scale structure, so alternative models may be tested by comparing their prediction of how structure grows to observations. The linear growth rate of structure, $f = d \ln(D)/d \ln(a)$, is the cosmological parameter that quantifies this growth, where $D$ is the linear growth function (describing how matter overdensities have evolved relative to early times), and $a$ is the scale factor, a dimensionless parameter that characterises the Universe's expansion. Modified gravity models commonly predict that the growth rate of structure should be a function of scale \citep[e.g.][]{Baker2014}, whereas general relativity predicts the growth rate to be scale-independent and related only to the matter density through $f=\Omega_m(z)^{0.55}$ \citep{Linder2005}. Consequently, a precise measurement of the growth rate of structure as a function of scale and redshift serves as an important test of our standard model and alternative modified gravity models.
	
	It is not only the distribution of galaxies that is useful in large-scale structure analyses, but also their motion in the form of peculiar velocities. A galaxy's peculiar velocity is the motion that is not due to cosmological expansion. Peculiar velocities have been shown to be a competitive cosmological probe when it comes to constraining the growth rate of structure \citep[e.g.][]{Koda2014,Howlett2017}. \cite{Johnson2014} used peculiar velocities to provide the first scale-dependent measurement of the growth rate of structure, which has since been used to test alternative parametrizations of gravity \citep{Johnson2016}.
	
	Direct peculiar velocities are particularly suited to probing large scales, making them highly complimentary to other cosmological probes, including redshift-space distortions (RSDs; see fig. 3 in \citealt{Johnson2014}, adapted from fig. 8 in \citealt{Jain2010}). However, their measurement is limited to low redshifts $(z\lesssim0.1)$ as their uncertainties grow with distance (see fig. 4 in \citealt{Scrimgeour2016} for an example). This complicates matters; the number of large-scale Fourier modes is limited when surveying a small volume of space, introducing a non-negligible sample variance contribution to the power spectrum. This means that the power spectrum for any observable tied to the fixed volume will have a fundamental lower bound on the uncertainty: it cannot be measured more precisely than the sample variance for a given scale. Since the growth rate of structure is inferred from the amplitude of the peculiar velocity power spectrum, it too is affected. 
	
	While this uncertainty floor exists, there are methods available to further lower the overall statistical uncertainty in the growth rate of structure; such improvements are essential if we wish to accurately test predictions from different cosmological models. Much work has been done in this area: for two observables that trace the same underlying matter distribution, it is possible to improve the uncertainty in the amplitudes of their power spectra by adding information about how the two tracers are correlated. Furthermore, neglecting stochasticity, the ratio of these two tracers is entirely independent of the underlying matter distribution, and will not contain a sample variance contribution \citep[e.g.][]{McDonald2009,Gil-Marin2010, Bernstein2011,Abramo2013}. The most common application of this effect for peculiar velocity studies is to include the galaxy overdensity field as a second tracer, which is proportional to the underlying matter overdensity field via the galaxy bias, $b$. It is then possible to relate the two fields by gravitational instability theory and constrain the ratio between the growth rate of structure and the galaxy bias, $\beta = f/b$, with high precision. This is done by observing both fields, and reconstructing one from the other for different values of $\beta$ \citep[e.g.][]{Pike2005,Davis2011,Carrick2015}. If enough is known about the galaxy bias, the growth rate of structure can then be estimated. However, the method does not lend itself well to measuring the scale dependence of the growth rate of structure, and (as far as we are aware) no attempt has been made to do so.
	
	In this study we propose a new way of analysing the relation between the peculiar velocity and galaxy overdensity fields: instead of modelling one field in terms of the other, we model the covariance between the two fields, which can be done analytically. This covariance can then be used in a maximum likelihood method to constrain our cosmological parameters of interest: the growth rate of structure, $f$, and either the galaxy bias, $b$, or the ratio of these, $\beta$. We note that the expression for the cross-covariance has been previously presented by \cite{Fisher1995} but has never been applied to data. Our work builds on previous analyses that measure the amplitude of the velocity divergence power spectrum by modelling the peculiar velocity auto-covariance \citep{Jaffe1995,Macaulay2012}. More specifically, we follow on from \cite{Johnson2014}, who obtained a scale-dependent constraint on the growth rate of structure from the peculiar velocity auto-covariance.
	
	The aim of this paper is to lay down the theoretical foundation for the cross-covariance between the peculiar velocity and galaxy overdensity fields, so as to constrain the growth rate of structure to higher precision than can be obtained from peculiar velocities alone. We validate our approach by applying it to data from the GiggleZ N-body simulation, before analysing data from the 6-degree Field Galaxy Survey (6dFGS), obtaining constraints for the growth rate of structure. This is the first application of this method, and this paper serves to illustrate its effect on cosmological constraints when used with current peculiar velocity and redshift data.
	
	We begin by introducing the data in Section \ref{sec:data_sims}, and then discuss the theory and methodology in Section \ref{sec:theory_methodology}. Results for the simulation data are presented in Section \ref{sec:simresults} followed by the 6dFGS analysis in Section \ref{sec:6dfgsresults}. We conclude with a summary of the paper in Section \ref{sec:summary}.
	%
	%
	\section{Data and Simulations}\label{sec:data_sims}
	This analysis requires measurements of the galaxy overdensity field, $\delta_g(\bm{x})$, and the peculiar velocity field, $v_p(\bm{x})$. Both can be extracted from redshift surveys, although peculiar velocities also require an estimate of the true distance to the galaxy. These fields will be used to constrain our model of the cross-correlation and we can then infer constraints on our cosmological parameters of interest. 
	
	\subsection{6dFGS}  \label{subsec:6dFdata}
	We choose to work with data from the 6-degree Field Galaxy Survey (6dFGS, \citealt{Jones2004, Jones2005, Jones2009}) as it contains the largest single collection of peculiar velocity measurements currently available. 6dFGS is a redshift survey selected from the 2-Micron All-Sky Survey (2MASS) that covers the entire southern sky except for 10 degrees around the Galactic plane. It can be broken into two samples: the redshift sample, 6dFGSz, and the peculiar velocity sample, 6dFGSv.
	
	As of the final data release, 6dFGSz contains 125,071 extragalactic redshifts with a median redshift of $z =0.053$. Our redshift sample for calculating the galaxy overdensity comes from the 6dFGS baryon acoustic oscillation analysis by \cite{Beutler2011}, who selected galaxies with magnitude $K\leq12.9$ in regions of the sky that had a completeness value of 60\% or greater, which yielded 75,117 galaxies. 
	
\begin{figure*}
		\includegraphics[width=\textwidth]{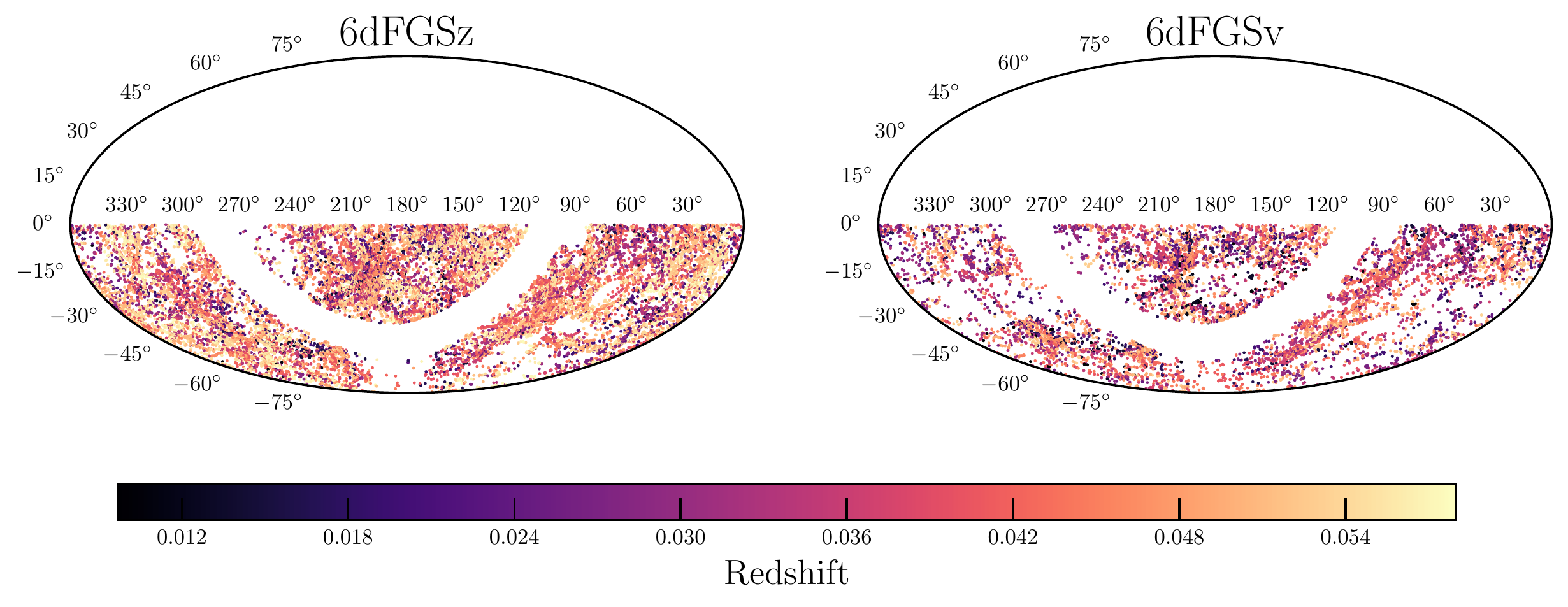}
		\caption{Distribution of the 6dFGSz and 6dFGSv samples, coloured by redshift. Both plots are equal-area Aitoff projections.}
		\label{fig:6dFdatadist}
	\end{figure*}

	In 6dFGS, peculiar velocities were obtained by applying the Fundamental Plane method to a high signal-to-noise subset of spectra. We use the 6dFGSv sample as defined by \cite{Springob2014}, which required signal-to-noise ratios greater than 5$\AA^{-1}$, and velocity dispersions greater than the resolution limit of the 6dF spectrograph ($\sigma_0 \geq$112\kms). This selection yielded 8,885 galaxies. Fig. \ref{fig:6dFdatadist} shows the sky and redshift distribution of both samples used in this work. 
	
	Our analysis proceeds by gridding the data into cubic cells. Following \cite{Johnson2014}, the peculiar velocity sample is gridded into cubes of length $L=10$\mpch. The redshift sample is gridded into cubes of length $L = 20$\mpch; we found that gridding at higher resolution did not affect our results so chose a larger gridding to reduce the dimension of our data vector. The gridding produced $N_{\delta} = 1036$ cells for 6dFGSz and $N_{v} = 2977$ cells for 6dFGSv. We now outline the steps involved in calculating the galaxy overdensity, and leave further discussion of the benefits and consequences of gridding to Section \ref{subsec:datamods}.
	
	We require a volume-limited sample to calculate the galaxy overdensity; this ensures that our sample is unbiased by magnitude and that the galaxy bias does not evolve with redshift. Data from 6dFGS is magnitude-limited, and we calculate the absolute magnitude using the \textit{k}-correction from \cite{Mannucci2001}. For simplicity, we limit our sample to the same volume as 6dFGSv, with a maximum redshift of $z=0.057$; this defines the absolute magnitude cut: galaxies need to be brighter than $M=-23.37$ to exceed the apparent magnitude threshold at all redshifts. After the cut, our redshift sample contains 20,796 galaxies. 
	
	The overdensity is related to the ratio between the number of galaxies in a cell, $N_{\rm cell}$, and the mean background number expected for that cell, $N_{\rm exp}$. Since the survey geometry and completeness affects the expected number of galaxies, we proceed by determining the selection function, assuming that the radial and angular components are separable. We obtain the radial selection function by taking a histogram of the sample and calculating the number density as a function of redshift, $n(z)$. There is a slight increase in number density with redshift, which can be explained by selection effects and our choice of \textit{k}-correction, so we fit a second order polynomial to obtain a functional form for $n(z)$. We use the angular selection function derived for 6dFGS by \cite{Jones2006}. We calculate the number of galaxies expected in a given grid cell and normalise this to the total number of galaxies in our sample. From this, we calculate the overdensity using $\delta_g = \tfrac{N_{\rm cell}}{N_{\rm exp}} -1$, and obtain the shot noise for each cell assuming Poisson statistics as $\sigma_{\delta_g} = 1/\sqrt{N_{\rm exp}}$. 
	
	The peculiar velocity measurement for each cell is the average of all velocities in that cell, and we add the observational uncertainties in quadrature. See \cite{Abate2008} for a discussion of the motivation behind this approach.
	
	Due to the distribution of galaxies in 6dFGS, there are cells that do not contain any galaxies. For example, a cell that covers the Galactic plane will be empty since the survey did not take any data in this region. For peculiar velocities, a cell can only be empty if we have no measurement there: the cell has no information to contribute to our analysis and can be safely removed. The overdensity cells are more complicated, since an empty cell provides information about the overdensity field as long as it is within the survey volume. Thus, we keep empty overdensity cells if they fall within the 6dFGS survey volume and exclude them otherwise.
	
	Finally, since our sample is at very low redshift, we take the effective redshift for our growth rate of structure measurement to be at $z=0$. 
	
	\subsection{Simulation} \label{subsec:simulationdata}
	We use simulated peculiar velocities and halo positions from the GiggleZ simulation \citep{Poole2014}. The volume of the simulation is 1 (\gpch)$^3$, and it contains $2160^3$ particles, each with a mass of $7.5\times10^9$ \msol. It was generated using GADGET-2 \citep{Springel2005}, and halos and subhaloes were identified with the \texttt{SUBFIND} algorithm \citep{Springel2001}. The fiducial cosmology for GiggleZ is a spatially-flat $\Lambda$CDM fit to the Wilkinson Microwave Anisotropy Probe (WMAP) five-year data (with the addition of baryon acoustic oscillation and supernova data), and the cosmological parameters are listed in Table \ref{tab:powerspectrummodels} in Section \ref{subsec:fidpowerspectrum}.
	
	Applying our method to simulated data allows us to check how well the approach recovers known input cosmological parameters, as well as the effects of introducing the cross-covariance in the absence of noise. We create a single approximate realisation of the 6dFGS survey from GiggleZ, constructing a hemisphere around an observer out to redshift $z=0.057$ (the maximum redshift of the 6dFGSv sample), corresponding to a radius of $\sim 150$\mpch. For our galaxy overdensity sample, we use haloes with subhalo masses between $10^{13}$\msol\ and $10^{13.5}$\msol. This limits the variation in galaxy bias for our sample, as galaxy bias is related to halo mass \citep[e.g.][]{Seljak2005}. Linear theory prescribes that all velocities should be fair tracers of the underlying matter distribution, so we use all velocities available in our chosen volume. We project the simulated velocities onto the line of sight to obtain the peculiar velocity, and then convert this to a logarithmic distance ratio, $\eta$, which is the observed quantity from the 6dFGS Fundamental Plane analysis (see Section \ref{subsec:datamods}).
	
	As with 6dFGS, we grid each sample to obtain a measurement of the galaxy overdensity and average peculiar velocity in each cell. We find that gridding both samples into cells of length $L=20$\mpch\, is sufficient for recovering the simulation cosmology and has the added advantage of improving the computation speed over a higher-resolution gridding (see Section \ref{subsec:datamods}). Since our samples from the simulation are not affected by a selection function, we can directly calculate the overdensity relative to the average number density for our sample. 
	
	We note that this realisation is not a true mock of 6dFGS as it does not contain observational errors for the peculiar velocities. This is desirable since it allows us to perform a more accurate test of our approach. The number densities are also different: we estimate that our GiggleZ sample has a galaxy number density of $n_{g}=1.7\times10^{-4}$ (\mpch)$^{-3}$ and peculiar velocity number density of $n_{ v}=1.0\times10^{-2}$(\mpch)$^{-3}$, whereas our 6dFGSz sample has a number density of $n_{ g}=3.6\times10^{-3}$(\mpch)$^{-3}$, and our 6dFGSv sample has a number density of $n_{ v}=1.5\times10^{-3}$(\mpch)$^{-3}$. This does not affect the ability of our GiggleZ sample validate our method, since we seek only to recover the input cosmological parameters.
	
	\section{Theory and Methodology}\label{sec:theory_methodology}
	It is physically intuitive that the overdensity field and peculiar velocity field are correlated: increasing the matter density at a given position will pull galaxies towards it though gravity. If we are to use this information to test cosmological models, we must formalise the theory for this correlation. This section provides a description of the theoretical constructs used and their applications.
	
\begin{figure}
		\includegraphics[width=\columnwidth]{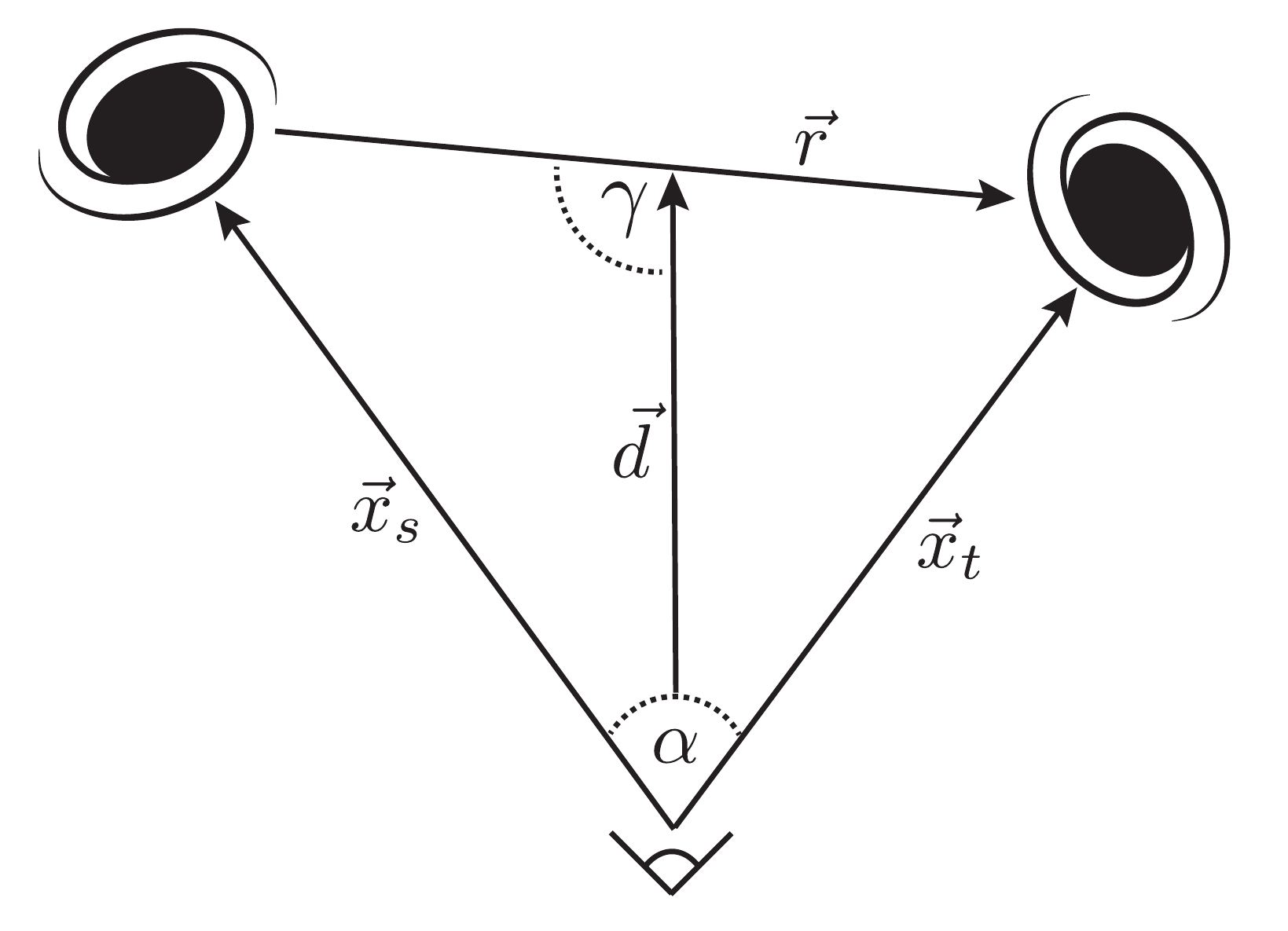}
		\caption{The covariance is always calculated between a pair of observations at arbitrary positions $\bm{x}_t$ and $\bm{x}_s$, separated by $\bm{r} = \bm{x}_t - \bm{x}_s$ and angle $\alpha$. When calculating the line of sight for a pair, we use the dot product $\bm{d}\cdot\bm{r}=\cos(\gamma)$, where $\bm{d} = \tfrac{1}{2}(\bm{x}_t + \bm{x}_s)$. All distances are measured in \mpch.}
		\label{fig:vectordef}
	\end{figure}

	Throughout this paper, we will use a set of definitions for relative positions and angles, which are shown in Fig. \ref{fig:vectordef}. All positions are given in units of \mpch, where the Hubble constant at redshift $z=0$ is $H_0 = 100$ \hubbleunit, and the reduced Hubble constant, $h$, is given in Table \ref{tab:powerspectrummodels} for each cosmology.
	
	\subsection{Likelihood approach} \label{subsec:likelihoodapproach}
	Our aim is to model the galaxy overdensity and peculiar velocity fields in a self-consistent way. For this, we expand on the approach presented by \cite{Johnson2014}, which was developed for the peculiar velocity field only. This relies on the construction of an appropriate likelihood function, which we detail now.
	
	We begin by assuming that the measured galaxy overdensities, $\bm{\delta}_g = (\delta_{g_1},\delta_{g_2},...\delta_{g_{N_\delta}})$, are correlated samples from an underlying multivariate Gaussian distribution, which has a mean of zero. We assume the same for the peculiar velocities, $\bm{v}_p = (v_{p_1},v_{p_2},...v_{p_{N_v}})$. These assumptions allow us to construct a theoretical likelihood function, which is the probability of observing the data given our model:
	\begin{align}
	\mathcal{L} &= p(\bm{\Delta}|\bm{\phi})  \\
	&= \frac{1}{\sqrt{(2\pi)^N |\mathbfss{C}(\bm{\phi})|}}\exp\left(-\frac{1}{2} \bm{\Delta}^T\mathbfss{C}(\bm{\phi})^{-1}\bm{\Delta}\right) \label{eq:likelihoodeq}
	\end{align}
	where $\bm{\Delta}$ is the data vector that contains the measured galaxy overdensities and peculiar velocities with length $N = N_v + N_\delta$, $\mathbfss{C}$ is the covariance matrix between each element of $\bm{\Delta}$, and $\bm{\phi}$ is a vector containing the parameters of the model we wish to constrain. In this approach, we construct a theoretical model for the covariance matrix in terms of $\bm{\phi}$. Since our model assumes that the mean values of the overdensity field and peculiar velocity field are zero, we do not explicitly include the subtraction of the mean from the data in the likelihood function.
	
	The covariance has the same dimension as the data, and the structure of the data vector and covariance is:
	\begin{eqnarray}
	\bm{\Delta} =  \begin{pmatrix}
	\bm{\delta}_g \\ \bm{v}_p
	\end{pmatrix}, \ \ \ \ \mathbfss{C} = \begin{pmatrix}
	\mathbfss{C}_{\delta_g \delta_g}  \ \mathbfss{C}_{\delta_g v_{p}} \\
	\mathbfss{C}_{v_{p} \delta_g} \ \mathbfss{C}_{v_{p} v_{p}}
	\end{pmatrix}. \label{eq:covmatrixstructure}
	\end{eqnarray}
	Each of the four submatrices in the covariance can be theoretically modelled in terms of the parameters of interest, which is discussed in Section \ref{subsec:modelcovariance}.
	
	The application of Bayes' theorem allows us to evaluate the probability of the model parameters given the data (the posterior) by modifying the underlying probability distribution of our model (the prior) through the ratio of the probability distribution of the data given the model (the likelihood) to the probability distribution of the data (the evidence). Mathematically:
	\begin{align}
	p(\bm{\phi}|\bm{\Delta}) = \frac{p(\bm{\phi})p(\bm{\Delta}|\bm{\phi})}{p(\bm{\Delta})}.
	\end{align}
	In order to evaluate this expression and find the values of the model parameters that maximise the posterior, we need to determine the prior and evidence (the likelihood is already described by Eq. \ref{eq:likelihoodeq}). When working with a fixed cosmological model, the evidence acts purely as a normalisation and does not affect our ability to determine which element of $\bm{\phi}$ provides the best fit to the data for our covariance model. If we take the prior on $\bm{\phi}$ to be uniform, this also becomes part of the normalisation and the maximum of the posterior can be determined purely from the maximum of the likelihood. We justify our use of a uniform prior in Section \ref{subsec:theoryforcosmologicalconstraints} when discussing the free parameters of our model.
	
	\subsection{Model covariance matrix} \label{subsec:modelcovariance}
	We now wish to model the four submatrices of the covariance as shown in Eq. \ref{eq:covmatrixstructure}. These matrices are the auto- and cross-correlation functions of the galaxy overdensity and peculiar velocity at two vector positions relative to the observer. These correlation functions can be determined from the auto- and cross-power spectra for the galaxy overdensity and velocity divergence fields for a given cosmological model.
	
	Assuming linear theory applies, the galaxy overdensity field is proportional to the matter overdensity field, $\delta_m$, and the velocity field is proportional to the velocity divergence field, $\theta$, in Fourier space:
	\begin{align}
	\delta_g(\bm{k}) &= b\delta_m(\bm{k}), \label{eq:deltag_deltam}\\
	\bm{v}(\bm{k}) &= -iaHf \frac{\bm{k}}{k^2} \theta(\bm{k}), \label{eq:v_theta}
	\end{align}
	where $b$ is the galaxy bias, $f$ is the growth rate of structure, $a$ is the scale factor, and $H$ is the Hubble constant in units of \hubbleunit. Since we are working in Fourier space, $\bm{k}$ is the wavevector, and its magnitude is given by the wavenumber, $k$, in units of \hmpc. We note that $\theta(\bm{k}) = \delta_m(\bm{k})$ in linear theory, but keep these separate for notational clarity. A complete derivation of Eq. \ref{eq:v_theta} can be found in Appendix \ref{sec:pv_fourier}.
	
	We have now met the first two free parameters of our model: the galaxy bias, $b$, and the growth rate of structure, $f$. We purposefully neglect the growth-rate information contained in redshift-space distortions (RSD) of the galaxy overdensity field, as we wish to test how direct peculiar velocities are correlated with the galaxy overdensity field. We discuss how this choice affects our results in Section \ref{sec:6dfgsresults}, and note that our method can be extended to include RSD, which we will present in future work (see Section \ref{subsec:futurework}).
	
	The correlation function is the ensemble average of the product of the first field with the complex conjugate of the second; for example: $C_{\delta_g \delta_g}(\bm{x}_1,\bm{x}_2)=\langle \delta_g(\bm{x}_1) \delta_g^*(\bm{x}_2) \rangle$. In Appendix \ref{sec:covariance_derivation} we use Eq. \ref{eq:deltag_deltam} and Eq. \ref{eq:v_theta} to derive the following expressions for the auto- and cross-covariance between two positions $\bm{x}_s$ and $\bm{x}_t$ separated by $\bm{r} = \bm{x}_t - \bm{x}_s$ and angle $\alpha$:
	\begin{align}
	C_{\delta_g \delta_g} (\bm{x}_s, \bm{x}_t) &= \frac{b^2}{2\pi^2}\int P_{mm}(k) W_{\delta_g \delta_g}(\bm{x}_s, \bm{x}_t, k) dk \label{eq:ddcov_final_maintext} \\
	C_{v_p v_p} (\bm{x}_s, \bm{x}_t) &= \frac{(aHf)^2}{2\pi^2} \int P_{\theta \theta}(k) W_{v_p v_p} (\bm{x}_s, \bm{x}_t ,k) dk \label{eq:vvcov_final_maintext} \\
	C_{\delta_g v_p} (\bm{x}_s, \bm{x}_t) &=\frac{-aHfb}{2\pi^2}\int P_{m\theta}(k) W_{\delta_g v_p}(\bm{x}_s, \bm{x}_t, k)  dk \label{eq:dvcov_final_maintext} \\
	C_{v_p \delta_g} (\bm{x}_s, \bm{x}_t) &=\frac{aHfb}{2\pi^2} \int P_{\theta m}(k) W_{v_p \delta_g} (\bm{x}_s, \bm{x}_t, k)  dk \label{eq:vdcov_final_maintext}
	\end{align}
	where
	\begin{align}
	W_{\delta_g \delta_g}(\bm{x}_s, \bm{x}_t, k) &= k^2 j_0(kr) \\
	W_{v_p v_p} (\bm{x}_s, \bm{x}_t ,k) &= \left( \frac{1}{3} \cos \alpha [j_0(kr) - 2j_2(kr) ] \right.  \nonumber \\
	&\qquad {} \left. + \frac{x_s x_t}{r^2} j_2(kr)\sin^2 \alpha \right) \\
	W_{\delta_g v_p}(\bm{x}_s, \bm{x}_t, k) &= k (\hat{\bm{x}_t}\cdot\hat{\bm{r}})j_1(kr) \\
	W_{v_p \delta_g} (\bm{x}_s, \bm{x}_t, k) &=k (\hat{\bm{x}_s}\cdot\hat{\bm{r}})j_1(kr)
	\end{align}
	are the analytic solutions to the angular integrals involved in the covariance derivations (and $j_n$ is the $n^{\rm th}$ spherical Bessel function). We note that the galaxy overdensity auto-covariance is a well-known result, and that the peculiar velocity auto-covariance has been previously derived in the literature \citep[e.g.][]{Ma2011}. To the best of our knowledge, our derivation of the theoretical cross-covariance between the galaxy overdensity and peculiar velocity fields is the first presented in the literature, although a similar expression for the cross-covariance appears in \cite{Fisher1995}.
	
\begin{figure}
		\includegraphics[width=\columnwidth]{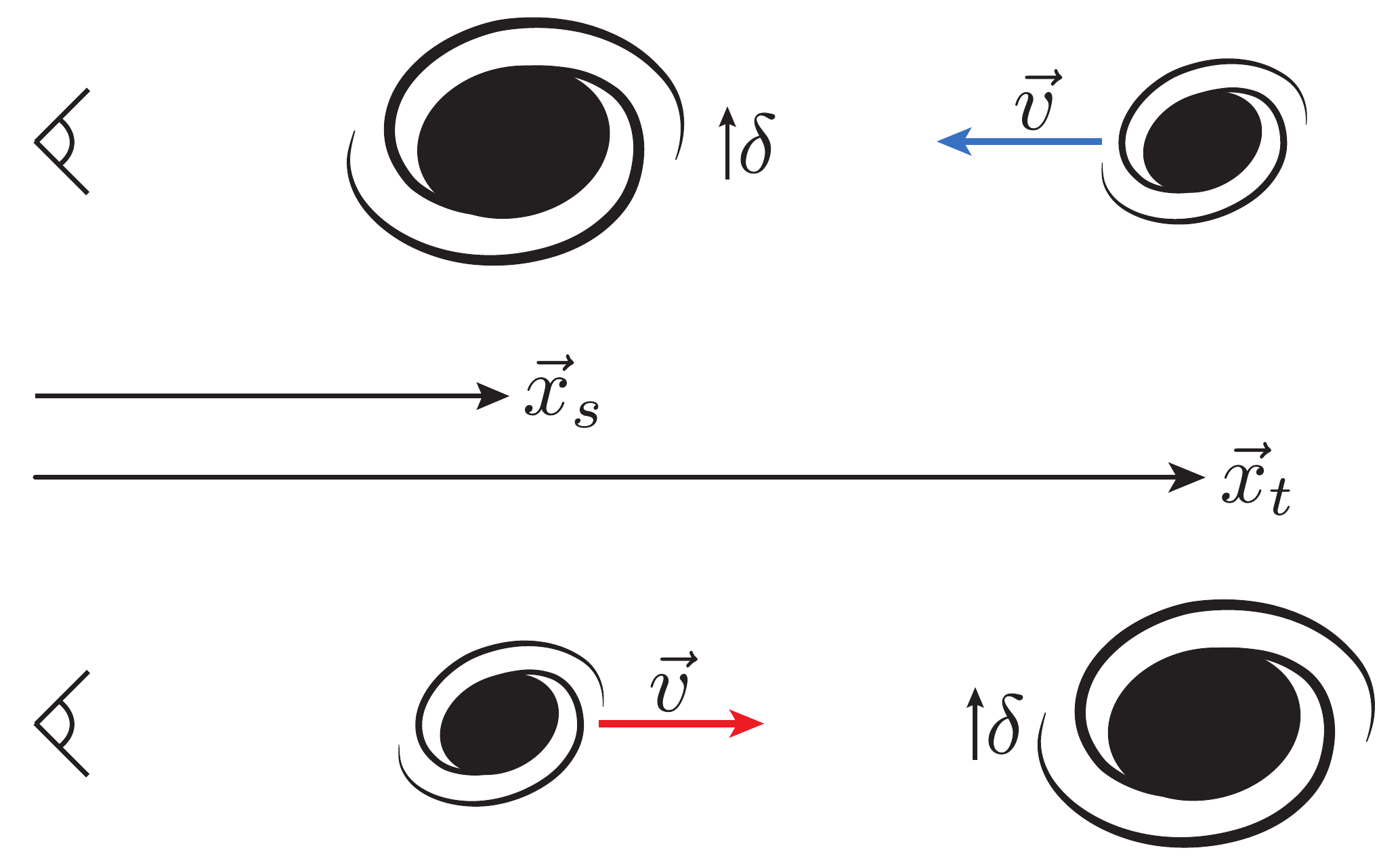}
		\caption{The sign of the covariance between a peculiar velocity and a galaxy overdensity is affected by their positions relative to the observer, as the direction of positive velocity is away from the observer. Here, we examine the effect of increasing the overdensity. In the top panel, the peculiar velocity increases towards the observer (becomes more negative), and in the bottom panel, it increases away from the observer (becomes more positive).}
		\label{fig:dvcorrelation}
	\end{figure}

	For the cross-covariance, it is especially important to note the distinction between the galaxy overdensity position and the peculiar velocity position, as it is always the peculiar velocity position that appears in the dot product with the vector separation of the two galaxies, $\bm{r}$. The difference in sign between Eq. \ref{eq:dvcov_final_maintext} and \ref{eq:vdcov_final_maintext} is illustrated by Fig. \ref{fig:dvcorrelation}, in which we show two example galaxies along the line of sight. If the galaxy overdensity increases, we expect the peculiar velocity as seen by the observer to increase in the direction of the overdensity. In the top panel, where the galaxy overdensity is closest to the observer, the peculiar velocity will increase towards the observer, giving negative peculiar velocity and leading us to expect a negative correlation (overdensity increases and peculiar velocity decreases). This situation corresponds to Eq. \ref{eq:dvcov_final_maintext} and the analytic covariance is negative, as expected. Switching the positions in the bottom panel, we expect a positive correlation, as the peculiar velocity will increase away from the observer towards the overdensity. In this case, Eq. \ref{eq:vdcov_final_maintext} applies and the analytic covariance is positive, as expected.
	
	Throughout this work, we have assumed that velocities are fair tracers of the matter overdensity field, implying that the velocity bias is unity. This is a fair assumption on linear scales, but becomes erroneous on small scales where the velocity bias introduces a scale-dependent modification to the amplitude of the peculiar velocity power spectrum. However, we consider a full systematic treatment of velocity bias to be beyond the scope of this work, and defer to \cite{Howlett2017} for a more detailed discussion of velocity bias and its impact on results using data from future surveys.
	
	\subsection{Modification of the covariance to account for peculiar velocity data} \label{subsec:modificationforvelocitydata}
	Although linear theory describes peculiar velocities as being drawn from a multivariate Gaussian with a mean of zero, this is not the case when observational errors are introduced. \cite{Springob2014} has shown that peculiar velocities from the 6-degree Field Galaxy Survey (6dFGS) have uncertainty distributions that are log-normal rather than Gaussian. We can account for this by updating our model for the covariance.
	
	For 6dFGS, peculiar velocities were obtained using the Fundamental Plane relation, which links the effective radius of an elliptical galaxy, its effective surface brightness, and its central velocity dispersion. It can be used to derive peculiar velocities because it acts as a redshift-independent distance estimator, which can break the degeneracy between the redshift from peculiar velocity, $z_{v_p} = v_p/c$, and the redshift from expansion, $z_{H}$, that make up the observed redshift of an object:
	\begin{equation}
	(1+z_{\rm obs}) = (1+z_{v_p})(1+z_H).
	\end{equation}
	As discussed by \cite{Magoulas2012}, the Fundamental Plane fit for each galaxy provides the probability distribution of the quantity $\log_{10}[R(z_{\rm obs})/R(z_{H})]$, which is the logarithmic ratio of a galaxy's observed effective radius (measured by the survey) to the effective radius due purely to Hubble flow (inferred from the Fundamental Plane fit). However, peculiar velocities are determined from the logarithmic ratio of the comoving distance inferred from the observed redshift, $D(z_{\rm obs})$, to the true comoving distance, $D(z_{H})$, which can be calculated from $R(z_{\rm obs})$ and $R(z_{H})$. \cite{Springob2014} derived the probability distribution for the logarithmic distance ratio $\eta \equiv \log_{10}[D(z_{\rm obs})/{D(z_H)]}$ for each galaxy, and \cite{Johnson2014} showed that the transformation between $p(\eta)$ and $p(v_p)$ is non-linear, resulting in a skewed distribution for $p(v_p)$ (see fig. 5 in that work).
	
	Since $\eta$ has a Gaussian distribution, we rewrite the likelihood in terms of this parameter. This requires a conversion factor from our modelled peculiar velocity (Eq. \ref{eq:v_theta}) to $\eta$. Such modelling has already been performed for how peculiar velocities affect supernova magnitudes \citep[e.g.][]{Hui2006}. The 6dFGS Fundamental Plane uses a different convention to type Ia supernovae, and we derive the conversion factor between peculiar velocity and the logarithmic distance ratio for 6dFGS in Appendix \ref{sec:vtoeta_theory}. This gives a single scaling factor which is a function of the observed redshift:
	\begin{align}
	\xi = \frac{1}{\ln(10)} \frac{(1+z_{\rm obs})}{D(z_{\rm obs})H(z_{\rm obs})},
	\end{align}
	where $H(z_{\rm obs})$ is Hubble's constant at the observed redshift, measured in \hubbleunit, and $D(z_{\rm obs})$ is the comoving distance corresponding to the observed redshift, measured in \mpch. Then,
	\begin{align}
	\eta = \xi v_p,
	\end{align}
	where $\eta$ is dimensionless, as expected for a logarithmic quantity. The equations for the logarithmic distance ratio auto-covariance and the cross-covariances then become:
	\begin{align}
	&C_{\eta \eta} (\bm{x}_s, \bm{x}_t)= \xi^2 C_{v_p v_p} (\bm{x}_s, \bm{x}_t) \\
	&C_{\delta_g \eta} (\bm{x}_s, \bm{x}_t) = \xi C_{\delta_g v_p} (\bm{x}_s, \bm{x}_t)\\
	&C_{\eta \delta_g} (\bm{x}_s, \bm{x}_t) = \xi C_{v_p \delta_g} (\bm{x}_s, \bm{x}_t)
	\end{align}
	
	\subsection{Inclusion of error terms} \label{subsec:inclusionoferror}
	We now discuss the inclusion of uncertainties; assuming that the errors on the data points are independent, these appear along the diagonal of the covariance matrix. For an observed value of $\eta$ at position $\bm{x}_i$, we include the observed error in $\eta$ from the Fundamental Plane, $\sigma_{\rm obs}(\bm{x}_i)$ (as detailed in \citealt{Springob2014}), and a stochastic velocity term to account for the breakdown of linear theory, $\sigma_{v}$. This acts as the third and final free parameter in our analysis, taking the same value for all diagonal entries of the peculiar velocity auto-covariance. The diagonal contribution to the logarithmic distance ratio auto-covariance will then be:
	\begin{align}
	&\sigma^2_{\eta \eta}(\bm{x}_s, \bm{x}_t) = \sigma^2_{\rm obs} (\bm{x}_s)\delta_{st} + \xi^2\sigma^2_{v}\delta_{st} \\
	&C_{\eta \eta} (\bm{x}_s, \bm{x}_t)= \xi^2 C_{v_p v_p} (\bm{x}_s, \bm{x}_t) + \sigma^2_{\eta \eta}(\bm{x}_s, \bm{x}_t),
	\end{align}
	where $\delta_{st}$ is the Kronecker delta.
	
	For an observed value of $\delta_g$ at position $\bm{x}_i$, we include the shot noise contribution, $\sigma_{\delta_g}(\bm{x}_i)$, assuming that the galaxy counts are drawn from a Poisson distribution, such that the shot noise variance is equal to the inverse of the average number count (see Section \ref{subsec:6dFdata}). The diagonal contribution to the galaxy overdensity auto-covariance will then be:
	\begin{align}
	&\sigma^2_{\delta_g \delta_g}(\bm{x}_s, \bm{x}_t) = \sigma^2_{g} (\bm{x}_s)\delta_{st} \\
	&C_{g g} (\bm{x}_s, \bm{x}_t)= C_{\delta_g \delta_g} (\bm{x}_s, \bm{x}_t) + \sigma^2_{\delta_g \delta_g}(\bm{x}_s, \bm{x}_t).
	\end{align}
	
	\subsection{Cosmological constraints} \label{subsec:theoryforcosmologicalconstraints}
	Our model parameters are the growth rate of structure, $f$, the galaxy bias, $b$, and the stochastic velocity, $\sigma_v$. We do not vary the underlying cosmological parameters in this study and instead use fixed fiducial power spectra $P_{mm}(k)$, $P_{m\theta}(k)$ and $P_{\theta \theta}(k)$ when calculating the covariances in Eq. \ref{eq:ddcov_final_maintext}-\ref{eq:vdcov_final_maintext}. We take uniform priors on the model parameters since we do not want to impose any restrictions on the relation between $f$ and $b$, and only impose the physically-motivated condition that none of our free parameters may be negative.
	
	Both the growth rate of structure and the galaxy bias scale the amplitude of the three power spectra, and are hence degenerate with another cosmological parameter, $\sigma_8$, which is the root-mean-square amplitude of linear matter fluctuations in spheres of 8\mpch. Since we have fixed the fiducial power spectra, we can divide out by the fiducial $\sigma_8$ value, and instead constrain the parameter combinations $f\sigma_8$ and $b\sigma_8$. For Eq. \ref{eq:ddcov_final_maintext}-\ref{eq:vdcov_final_maintext}, we substitute:
	\begin{align}
	f \rightarrow f\sigma_8/\sigma_{8}^{\rm fid}, \\
	b \rightarrow b\sigma_8/\sigma_{8}^{\rm fid}.
	\end{align}
	
	Since this is the first use of this approach to constrain cosmological parameters, we wish to examine how including the cross-covariance affects our constraints compared to only using the auto-covariance pieces. To do this, we consider two tests:
	\begin{enumerate}
		\item Assume the galaxy overdensity and peculiar velocity fields are uncorrelated, setting $\mathbfss{C}_{\delta_g v_p} = \mathbfss{C}_{v_p \delta_g} = \mathbfss{0}$.
		\item Include the cross-correlation, calculating $\mathbfss{C}_{\delta_g v_p}$ and $\mathbfss{C}_{v_p \delta_g}$.
	\end{enumerate}
	Comparing the constraints from these tests allows us to examine not only the effect of including the extra information, but whether the data supports a cross-correlation. In particular, we note that in the first test, $f\sigma_8$ is only constrained by the peculiar velocity auto-covariance, and $b\sigma_8$ is only constrained by the galaxy overdensity auto-covariance, whereas the second test also constrains their product, $fb\sigma_8^2$.
	
	Forecasts for cosmological surveys, which use the Fisher matrix formalism, have shown that constraints on cosmological parameters improve when the cross-covariance is included \citep[e.g.][]{Howlett2017}. While the constraint on the ratio of these parameters is improved due to the correlated sample variance of the fields, the individual parameters should also see improved constraints from needing to satisfy the extra relationship imposed by the cross-covariance. \cite{Abramo2013} demonstrated this by diagonalising the Fisher matrix for correlated data. 
	
	Traditionally, works that compare peculiar velocity and galaxy overdensity measurements tend to constrain the redshift-space distortion parameter, $\beta$, as opposed to the galaxy bias, where $\beta=f/b$ \citep{Pike2005,Davis2011,Carrick2015}. This comes from the relationship between the peculiar velocity field and the matter overdensity field described by gravitational instability theory \citep{Peebles1976}:
	\begin{align}
	\bm{v}(\bm{r}) &= \frac{aHf}{4\pi} \int d^3\bm{r}' \delta_m(\bm{r}')\frac{\bm{r}'-\bm{r}}{|\bm{r}'-\bm{r}|^3} \\
	&= \frac{aH\beta}{4\pi} \int d^3\bm{r}' \delta_g(\bm{r}')\frac{\bm{r}'-\bm{r}}{|\bm{r}'-\bm{r}|^3}. \label{eq:gravinst}
	\end{align}
	This applies to galaxy distributions that have been corrected for redshift-space distortions such that $\delta_g$ is in real-space. In our analysis, we have absorbed redshift-space distortions into the galaxy bias such that we are effectively fitting for the redshift-space galaxy bias, $b_s$, in terms of the real-space galaxy bias, $b_r$:
	\begin{align}
	b_s^2 &= b_r^2 \left(1+\frac{2}{3}\beta + \frac{1}{5}\beta^2 \right) \\
	&= f^2 \left(\frac{1}{5} + \frac{2}{3\beta} + \frac{1}{\beta^2} \right), \label{eq:realspacebeta}
	\end{align}
	where $b_s$ is boosted by the total contribution of redshift-space distortions to the power spectrum \citep{Kaiser1987}. This allows us to parametrize Eq. \ref{eq:ddcov_final_maintext}-\ref{eq:vdcov_final_maintext} in terms of $f$ and $\beta$. We will apply this parametrization to our 6dFGS analysis, but will continue to use $f$ and $b_r$ for the simulation data, since these are in real space.
	
	\subsection{Generation of the fiducial matter power spectrum} \label{subsec:fidpowerspectrum}
	The fiducial power spectra required for the covariances are computed prior to commencing the likelihood runs. We generated the velocity divergence power spectrum, $P_{\theta \theta}$, and the cross power spectrum, $P_{m\theta} = P_{\theta m}$, with \texttt{velMPTbreeze}. This is an extension to \texttt{MPTbreeze} (\citealt{Crocce2012}) and uses two loop multi-propagators to generate the power spectra. While \texttt{velMPTbreeze} also provides the matter power spectrum, $P_{mm}(k)$, we instead generate the non-linear matter power spectrum from the Code for Anisotropies in the Microwave Background (\texttt{CAMB}, \citealt{Lewis2000}; \citealt{Lewis2011}), which utilises non-linear corrections from Halofit. This is because the galaxy overdensity auto-covariance is influenced by scales beyond the linear regime (see section \ref{subsec:integrationbounds}), so we require a more accurate determination of the matter power spectrum.
	
	For our simulation tests, the power spectra are generated from the same cosmological parameters as used in GiggleZ, which allows us to check whether we recover the simulation input values. For the 6dFGS analysis, we use power spectra generated from the median values of the 6-parameter base $\Lambda$CDM model for the Planck 2015 TT-lowP data (\citealt{PlanckCollaboration2015}). To check how our results are affected by this choice, we also use power spectra generated from the median values of the 6-parameter base $\Lambda$CDM model for the Wilkinson Microwave Anisotropy Probe (WMAP) five-year data (\citealt{Komatsu2009}). We list these parameters in Table \ref{tab:powerspectrummodels}.
	\begin{table}
		\centering
		\caption{Cosmological parameters for the three cosmologies used in this analysis. The top section shows the 6 base parameters for standard $\Lambda$CDM: physical baryon density; physical dark matter density; reduced Hubble constant; scalar spectral index; scalar amplitude (with pivot point $k_0=0.002$\hmpc \ for GiggleZ and WMAP, and $k_0=0.05$\hmpc \ for Planck); and reionization optical depth.The bottom section shows $\sigma_8$, which is a derived parameter.}
		\label{tab:powerspectrummodels}
		\begin{tabular}{llll} 
			\hline
			& GiggleZ & Planck & WMAP\\
			\hline
			$\Omega_bh^2$ & 0.02267 & 0.02222 & 0.02273\\
			$\Omega_ch^2$ & 0.1131 & 0.1197 & 0.1099\\
			$h$ & 0.705 & 0.6731 & 0.719\\
			$n_s$ & 0.960 & 0.9655 & 0.963\\
			$A_s$ & 2.445$\times10^{-9}$ & 2.195$\times10^{-9}$ & 2.41$\times10^{-9}$\\
			$\tau$ & 0.084 & 0.078 & 0.087\\
			\hline
			$\sigma_8^{\rm fid}$ & 0.812 & 0.8417 & 0.7931\\
			\hline
		\end{tabular}
	\end{table}
	
	\subsection{Gridding of data} \label{subsec:datamods}
	In our analysis we grid the data into cells, and determine the covariance between cells. Gridding allows us to:
	\begin{itemize}
		\item Calculate the overdensity in each cell
		\item Smooth over non-linear effects
		\item Reduce the computation time by lowering the dimensionality of the covariance matrix and data vector.
	\end{itemize}
	See Section \ref{subsec:6dFdata} for details of how the galaxy overdensity and peculiar velocity is calculated for each cell.
	
	The second two points have been discussed by \cite{Abate2008} and \cite{Johnson2014}. Most importantly, the gridding has two effects on the modelling. The first is that smoothing samples onto a grid will reduce small-scale power, which we account for by multiplying the power spectra by a window function. We use a cubic gridding approach, where each grid cell has length $L$ in \mpch, corresponding to a sinc function in Fourier space. In three dimensions, the window function becomes:
	\begin{equation}
	\Gamma(k,L) = \left\langle \frac{8}{L^3} \frac{\sin  \left (k_x \tfrac{L}{2}\right )}{k_x} \frac{\sin  \left (k_y \tfrac{L}{2}\right )}{k_y} \frac{\sin  \left (k_z \tfrac{L}{2}\right )}{k_z} \right\rangle_{\bm{k} \in k},
	\end{equation}
	where the average is applied to all $\bm{k}$ vectors that have magnitude $k$. Since we may use different gridding sizes for peculiar velocities and overdensities, we define $\Gamma_{\delta_g}(k) = \Gamma(k,L_{\delta_g})$ and $\Gamma_{v_p}(k) = \Gamma(k,L_{v_p})$.
	Our modified covariance equations then become:
	\begin{align}
	C'_{\delta_g \delta_g}(\bm{x}_s, \bm{x}_t) &= \frac{b^2}{2\pi^2}\int P_{mm}(k) W_{\delta_g \delta_g}(\bm{x}_s, \bm{x}_t, k) \Gamma^2_{\delta_g}(k) dk \label{eq:ddcov_gridded_maintext} \\
	C'_{v_p v_p} (\bm{x}_s, \bm{x}_t) &= \frac{(aHf)^2}{2\pi^2} \int P_{\theta \theta}(k) W_{v_p v_p} (\bm{x}_s, \bm{x}_t ,k) \Gamma^2_{v_p}(k) dk \label{eq:vvcov_gridded_maintext} \\
	C'_{\delta_g v_p} (\bm{x}_s, \bm{x}_t) &= \frac{-aHfb}{2\pi^2}\int P_{m\theta}(k) W_{\delta_g v_p}(\bm{x}_s, \bm{x}_t, k)  \nonumber \\
	&\qquad \qquad \qquad  {} \Gamma_{\delta_g}(k)  \Gamma_{v_p}(k) dk \label{eq:dvcov_gridded_maintext} \\
	C'_{v_p \delta_g} (\bm{x}_s, \bm{x}_t) &= \frac{aHfb}{2\pi^2}\int P_{\theta m}(k) W_{v_p \delta_g}(\bm{x}_s, \bm{x}_t, k)  \nonumber \\
	&\qquad \qquad \qquad  {} \Gamma_{v_p}(k)  \Gamma_{\delta_g}(k) dk \label{eq:vdcov_gridded_maintext}.
	\end{align}
	After gridding, the position of the galaxy overdensity and average peculiar velocity for each cell is set to be the centre of that cell.
	
	The second effect is that a correction must be applied to the diagonal elements of the peculiar velocity auto-covariance to model the varying shot noise involved in taking the average of $N_i$ peculiar velocities in each cell. \cite{Abate2008} proposed the following correction:
	\begin{align}
	C_{v_p v_p}^{\rm corr}(\bm{x}_i,\bm{x}_j) &= C'_{v_p v_p}(\bm{x}_i,\bm{x}_j) \nonumber \\
	&\qquad {} + \frac{C_{v_p v_p}(\bm{x}_i,\bm{x}_j) - C'_{v_p v_p}(\bm{x}_i,\bm{x}_j)}{N_i} \delta_{ij},
	\end{align}
	which we adopt in our analysis.
	
	\subsection{Integration bounds}\label{subsec:integrationbounds}
	In theory the integral over $k$ in Eq. \ref{eq:ddcov_gridded_maintext} - \ref{eq:vdcov_gridded_maintext} must be evaluated from $k=0$ to $k=\infty$, which is not practical. As such, we must pick limits for the integral. The lower limit is dictated by the largest-scale mode that the fields are sensitive to. For the simulated data, this scale is the side-length of the simulation box (1 $h^{-1}$Gpc), which corresponds to $k_{\rm min}=0.006$\hmpc. The 6dFGS data may be influenced by modes larger than the survey volume, so there is no explicit restriction on the value of the minimum wavenumber. We set $k_{\rm min}=0.0025$\hmpc, similar to values used in the peculiar velocity auto-covariance studies conducted by \cite{Johnson2014} and \cite{Macaulay2012}. The upper limit is dictated by our ability to model the small, non-linear scales. \cite{Johnson2014} found that setting $k_{\rm max} = 0.15$\hmpc\ for the peculiar velocity auto-covariance provided a good compromise between accurate recovery of fiducial parameters and constraining power (see fig. 8 of that paper). We adopt this value for our peculiar velocity auto-covariance and find that it is also suitable for the cross-covariance (for both data sets). 
	
	For the galaxy overdensity auto-covariance, we find that there is a significant contribution to the integral for $k>0.15$\hmpc. We therefore add a second integral, which we refer to as the additional integral. This integral has a fixed galaxy bias, and acts to increase the value of the covariance without needing to provide an advanced model of the bias on non-linear scales. The integral for the galaxy overdensity auto-covariance then becomes:
	\begin{align}
	C'_{\delta_g \delta_g} (\bm{x}_s,\bm{x}_t) &= b_{\rm fit}^2\int_{k_{\rm min}}^{k_{\rm max}} f(k, \bm{x}_s,\bm{x}_t) dk \nonumber \\
	&\qquad+ b_{\rm add}^2\int_{k_{\rm max}}^{k_{\rm add}}f(k, \bm{x}_s,\bm{x}_t) dk \label{eq:fidintegral}
	\end{align}
	where $f(k)$ represents the integrand in Eq. \ref{eq:ddcov_gridded_maintext}. We choose $k_{\rm add} = 1.0$\hmpc, as this is where the gridding window function $\Gamma^2_{\delta_g}(k)$ becomes close to zero. In Section \ref{subsec:gigglezadditionalbias} we show that the additional integral is required for the simulation data and determine the best-fitting value for $b_{\rm add}$. We repeat this analysis for 6dFGS in Section \ref{subsec:6df_addbiasresults}, and show how the value of $b_{\rm add}$ affects the constraint on $f\sigma_{8}$.
	
	\subsection{Evaluating the likelihood function}
	We can now construct the final covariance. The data vector and covariance now have the following structure:
	\begin{align}
	\bm{\Delta} =  \begin{pmatrix}
	\bm{\delta}_g \\ \bm{\eta}
	\end{pmatrix}, \ \ \ \ \mathbfss{C} = \begin{pmatrix}
	\mathbfss{C}_{g g}  \ \mathbfss{C}_{g \eta} \\
	\mathbfss{C}_{\eta g} \ \mathbfss{C}_{\eta \eta}
	\end{pmatrix}, \label{eq:covmatrixstructure_final}
	\end{align}
	where the data vector, $\bm{\Delta}$, contains the galaxy overdensity and logarithmic distance ratio measurements for each grid cell, and the covariance components have the form:
	\begin{align}
	\mathbfss{C}_{gg} &= \mathbfss{C}'_{\delta_g \delta_g} + \bm{\sigma}^2_{\delta_g \delta_g} \mathbfss{I} \\
	\mathbfss{C}_{\eta \eta} &= \xi^2 \mathbfss{C}^{\rm corr}_{v_p v_p} + \bm{\sigma}^2_{\eta \eta} \mathbfss{I} \\
	\mathbfss{C}_{g\eta} &= \xi \mathbf{C}'_{\delta_g v_p} \\
	\mathbfss{C}_{\eta g} &= \xi \mathbf{C}'_{v_p \delta_g}.
	\end{align}
	We wish to constrain the free parameters $f\sigma_8$, $b\sigma_8$ and $\sigma_v$, and do this by evaluating the likelihood equation (Eq. \ref{eq:likelihoodeq}) for a gridded parameter space. We invert the covariance matrix by applying the GNU Science Library Cholesky solver to the equation $\mathbfss{C}\bm{\kappa} = \bm{\Delta}$, which yields $\bm{\kappa} = \mathbfss{C}^{-1}\bm{\Delta}$. The exponent of the likelihood equation is obtained by multiplying this by $-\tfrac{1}{2}\bm{\Delta}^T$. We analyse our results using the publicly available \texttt{ChainConsumer} \texttt{Python} package \citep{Hinton2016}, which provides parameter constraints for gridded likelihood evaluations.
	
\section{Simulation results} \label{sec:simresults}
\subsection{Parameter constraints} \label{subsec:simparamconstraints}
In this section we use data from an N-body simulation to validate our method and examine how the addition of the cross-covariance affects the analysis (discussed in Section \ref{subsec:theoryforcosmologicalconstraints}). We constructed an approximate realisation of the 6-degree Field Galaxy Survey from GiggleZ as discussed in Section \ref{subsec:simulationdata}. Since the simulation data is in real space, all references to galaxy bias within this section refer to the real-space galaxy bias, $b_r$.

We expect to recover the standard model prediction of the growth rate of structure, which is calculated from $f(z=0) = \Omega_m(z=0)^{0.55}$. We obtain the best-fitting value for the additional bias, $b_{\rm add}$, by fitting for this along with the standard bias, $b_{\rm fit}$, using only the galaxy overdensity data (see Section \ref{subsec:gigglezadditionalbias}). We find $b_{\rm fit} = 1.59$ and $b_{\rm add} = 1.61$ for our GiggleZ galaxy overdensity sample. For the fiducial $\sigma_8$ value of GiggleZ (listed in Table \ref{tab:powerspectrummodels}), we expect to recover $f\sigma_8 = 0.398$ and $b_{\rm fit}\sigma_8=1.29$. We do not set an expected value for $\sigma_v$, as this only serves as a nuisance parameter in this test. Applying a gridding length of $L = 20$\mpch\ for the galaxy overdensity and peculiar velocity data is sufficient to recover the input parameters. We also tested $k_{\rm max}$ values of $0.10$ and $0.20$\hmpc, and recover the input parameters at the 2$\sigma$ level in both cases.

\begin{figure*}
	\centering
	\begin{subfigure}{0.5\textwidth}
		\includegraphics[width=\columnwidth]{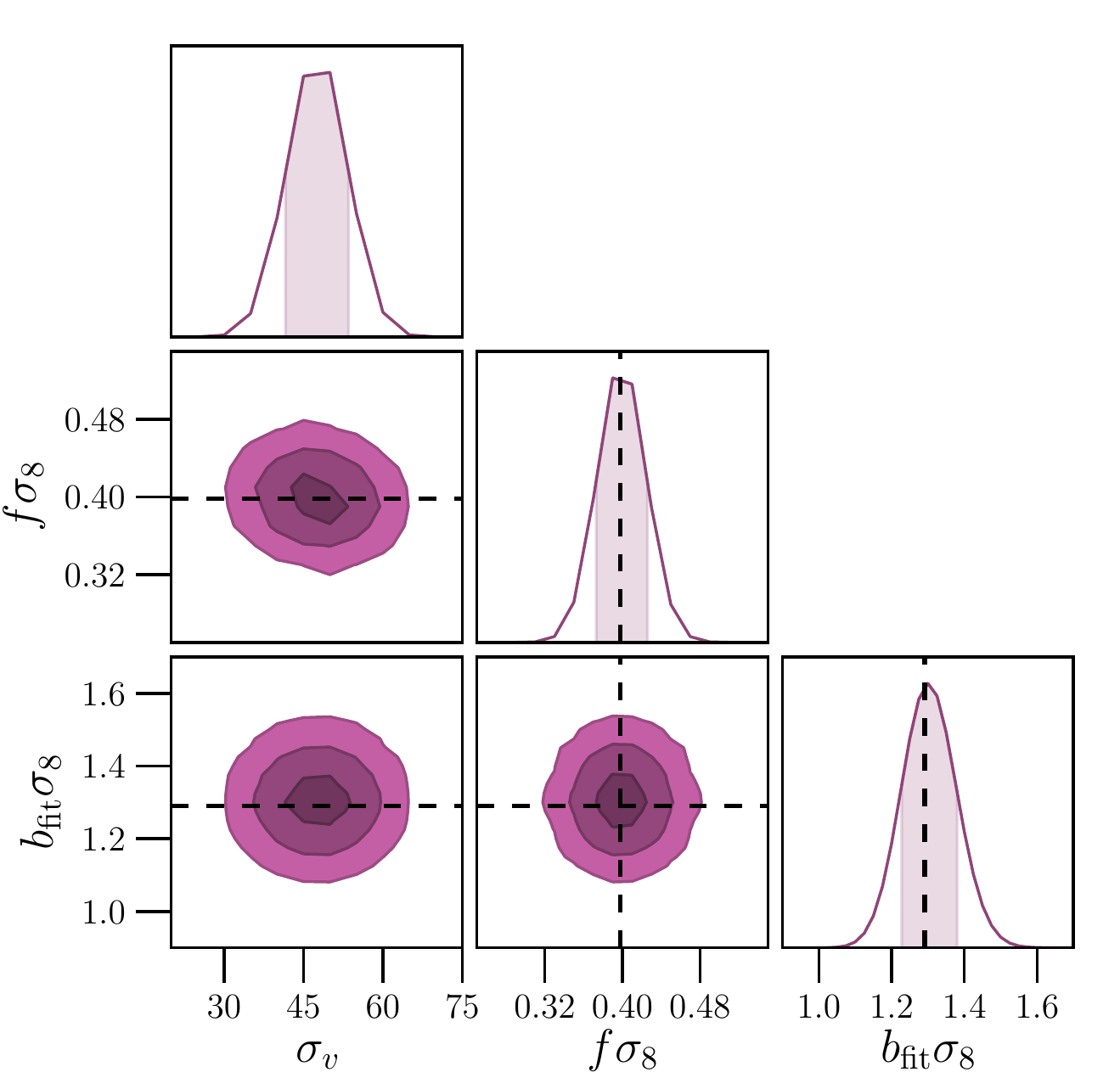}
		\caption{Cross-covariance excluded} \label{fig:gigglez_sim}
	\end{subfigure}%
	~ 
	\begin{subfigure}{0.5\textwidth}
		\includegraphics[width=\columnwidth]{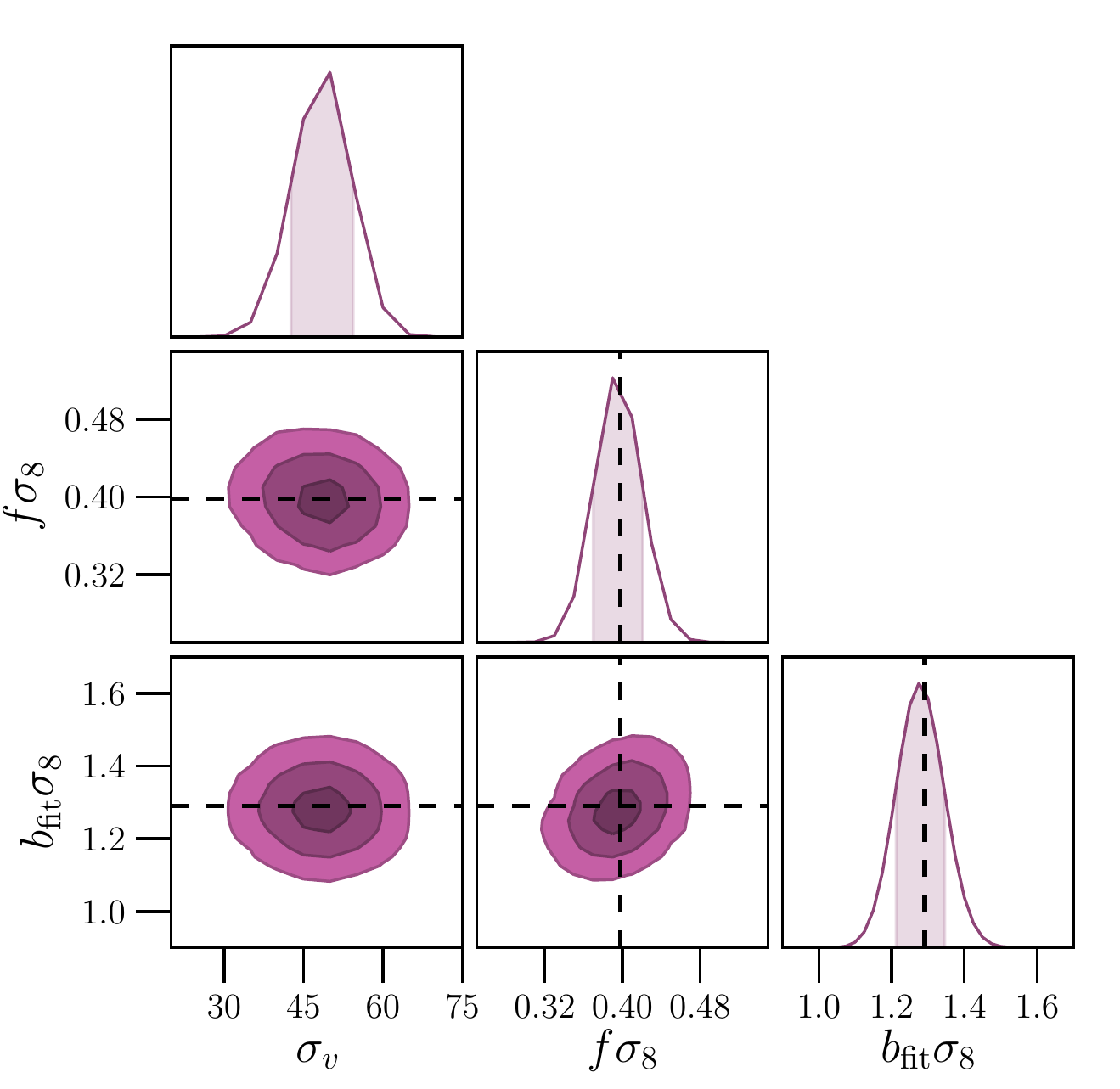}
		\caption{Cross-covariance included} \label{fig:gigglez_joint}
	\end{subfigure}
	\caption{The posteriors of our free parameters, $\sigma_v$ (\kms), $f\sigma_8$ and $b_{\rm fit}\sigma_8$, with shaded 68\% credible intervals, when fitting to the GiggleZ sample. The black dashed lines show the expected $f\sigma_8$ and $b_{\rm fit}\sigma_8$ values assuming the fiducial GiggleZ cosmology (Table \ref{tab:powerspectrummodels}) and an additional bias of $b_{\rm add} = 1.60$. Our analysis pipeline successfully recovers these values.}
\end{figure*}

The posteriors for our free parameters when assuming no cross-covariance ($\mathbfss{C}_{g \eta} = \mathbfss{C}_{\eta g} = \mathbfss{0}$) are shown in Fig. \ref{fig:gigglez_sim}, and the posteriors when including the cross-covariance are shown in Fig. \ref{fig:gigglez_joint}. The maximum likelihood and median (with 68\% credible intervals) are given in Table \ref{tab:GiggleZresults}. We obtain a 7\% measurement of $f\sigma_8$ and a 5\% measurement of $b_{\rm fit}\sigma_8$. Both tests recover the input parameters at the 1$\sigma$ level with a reasonable $\chi^2/\rm dof$, validating our analysis pipeline.

\begin{table}
	\centering 
	\caption{Maximum likelihood (ML) and median values (with 68\% credible intervals) of our free parameters for the fit to the GiggleZ sample. We include the $\chi^2$ value and $\chi^2$/dof statistic for the maximum likelihood.} \label{tab:GiggleZresults}
	\begin{tabular}{lcccc} 
		\hline
		&\multicolumn{2}{c}{No cross-covariance}&\multicolumn{2}{c}{Cross-covariance} \\
		& ML & Median & ML & Median \\
		\hline
		$\sigma_v$ (\kms) & 50.0 & $47.6\pm6.0$ & 50.0 & $48.6 ^{+5.8}_{-6.0}$ \\
		$f\sigma_8$ & 0.390& $0.399^{+0.027}_{-0.026} $ & 0.390 & $0.395\pm0.026$ \\
		$b_{\rm fit}\sigma_8$ &1.3&$1.303^{+0.077}_{-0.075}$ &1.275& $1.279^{+0.068}_{-0.066}$ \\
		$\chi^2$ & 2480.52 & - & 2484.09 & - \\
		$\chi^2/{\rm dof}$ & 0.99 & - & 0.99 & - \\
		\hline        
	\end{tabular}
\end{table}

The results of our GiggleZ test provide useful insight into our method. When setting the cross-covariance to zero, we would expect the $b_{\rm fit}\sigma_8$-$f\sigma_{8}$ contour to be uncorrelated, since $\mathbfss{C}_{gg}$ only constrains $b_{\rm fit}\sigma_8$ and $\mathbfss{C}_{\eta \eta}$ only constrains $f\sigma_8$. We calculated the Pearson correlation coefficient from our gridded likelihood result, finding $\rho = -6.3\times10^{-9}$, indicating an almost non-existent correlation. This is represented visually in Fig. \ref{fig:gigglez_sim}, since the contour forms an ellipse that is aligned with both axes.

Consequently, if a correlation exists between galaxy overdensity and peculiar velocity, then including the cross-correlation should change the tilt of the ellipse. We see in Fig. \ref{fig:gigglez_joint} that this is the case: the contour has become correlated and exhibits a positive slope. This highlights that the data are intrinsically correlated and that the cross-covariance has the power to constrain the relationship between $b_{\rm fit}\sigma_8$ and $f\sigma_8$. Here, we find a Pearson correlation coefficient of $\rho = 0.48$, indicating a moderate correlation, which is consistent with our visual interpretation.

Multiple works have indicated that we expect to see an improvement in constraints when including cross-covariances (see Section \ref{subsec:theoryforcosmologicalconstraints}). We find that this is the case for our correlated parameters, $b_{\rm fit}\sigma_8$ and $f\sigma_8$, although the improvement is greater for the galaxy bias. This may be in part because we do not include any observational uncertainty for the peculiar velocity auto-covariance in order to test the recovery of the fiducial model more precisely. Coupling this with the fact that the number density for our peculiar velocity sample ($n_{ v}=1.0\times10^{-2}$ (\mpch)$^{-3}$) is higher than that for the galaxy overdensity sample ($n_{g}=1.7\times10^{-4}$ (\mpch)$^{-3}$), we can also explain why the fractional uncertainty is lower for $f\sigma_8$ than for $b\sigma_8$. If a parameter is already so well determined from having a high number density and low uncertainties, the cross-covariance may contribute less information. We expect this to change when working with the 6dFGS data.

\subsection{Additional bias} \label{subsec:gigglezadditionalbias}
\begin{figure}
	\includegraphics[width=\columnwidth]{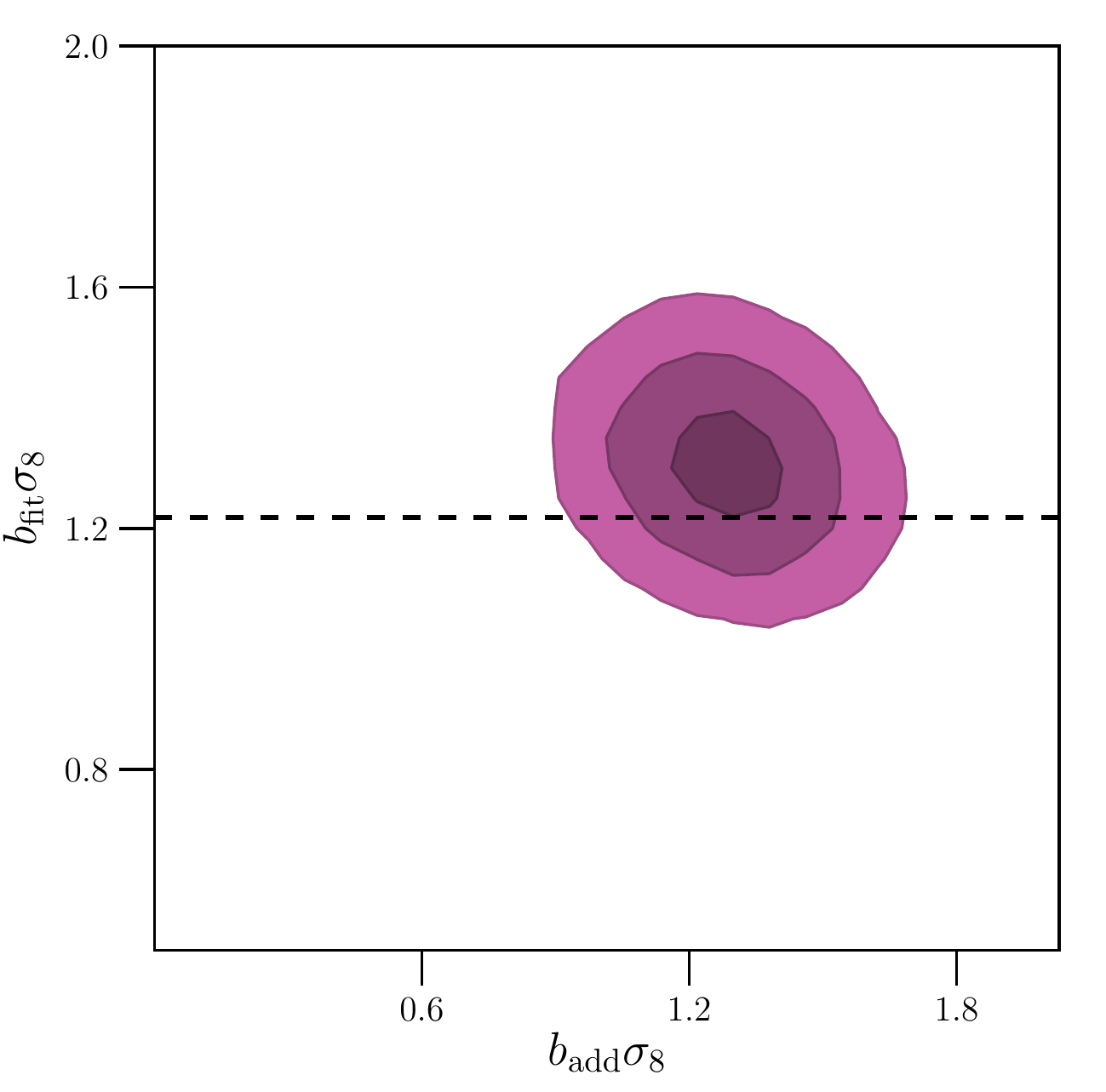}
	\caption{68, 95 and 99\% credible regions of $b_{\rm fit}\sigma_8$ and $b_{\rm add}\sigma_8$ when fitting the GiggleZ galaxy overdensity sample. The black dashed line represents the $b_{\rm fit}\sigma_8$ we expect to recover given the mass range of our GiggleZ galaxy overdensity sample.}
	\label{fig:gigglez_addbiastest}
\end{figure}

In this section, we show that the GiggleZ data requires the additional integral discussed in Section \ref{subsec:integrationbounds}, and determine the best-fitting value for $b_{\rm add}$ for our galaxy overdensity sample. We do this by evaluating the likelihood using only $\mathbfss{C}_{\delta_g \delta_g}(\bm{\phi})$: 
\begin{align}
\mathcal{L}_{gg} &= \frac{1}{\sqrt{(2\pi)^{N_\delta} |\mathbfss{C}_{\delta_g\delta_g}(\bm{\phi})|}}\exp\left(-\frac{1}{2} {\bm{\delta}_g}^T\mathbfss{C}_{\delta_g \delta_g}(\bm{\phi})^{-1}\bm{\delta}_{g}\right) \label{eq:likelihoodeq_galaxyonly}
\end{align}
where $\bm{\phi} = (b_{\rm fit}\sigma_8, b_{\rm add}\sigma_{8})$. We allow the $b_{\rm add}\sigma_8$ parameter to vary over [0,2], where a value of 0 indicates that the additional integral is not required by the data when constructing the covariance. The results are given in Fig. \ref{fig:gigglez_addbiastest}. 

We find median values of $b_{\rm fit}\sigma_8 = 1.29$ and $b_{\rm add}\sigma_8 =1.306$, with a $\chi^2$/dof of 1.01. $b_{\rm add} = 0$ is strongly disfavoured, with $\Delta\chi^2 = 241.27$. We take this as sufficient evidence for including the additional integral in our analysis. The case where $b_{\rm add} = b_{\rm fit}$ is within 1$\sigma$ of our median, which might suggest that the covariance needs to be fitted beyond our chosen $k_{\rm max}$ of 0.15 \hmpc. However, we do not wish to include non-linear information in our likelihood analysis, so fix the value of $b_{\rm add}$ rather than extending the value of $k_{\rm max}$. Our fitted bias is consistent with the power spectrum fits to GiggleZ data by \cite{Koda2014}, which suggest a bias value of $b\sigma_8 = 1.2$ for the mass range of our sample. 

\section{Data results} \label{sec:6dfgsresults}

\subsection{Parameter constraints} \label{subsec:6dfparamconstraints}
We now move to using the ($f\sigma_8$, $\beta$, $\sigma_v$) parametrization discussed in Section \ref{subsec:theoryforcosmologicalconstraints}. Our fiducial cosmology is the Planck 2015 base-set of parameters for $\Lambda$CDM, so we expect to recover $f\sigma_8 = 0.446$. The observed data is in redshift space, so all references to the galaxy bias within this section refer to the redshift-space bias, $b_s$. The best-fitting biases for our galaxy overdensity sample are $b_{\rm fit}\sigma_8 = 1.379$ and $b_{\rm add}\sigma_{8} = 1.50$ (see Section \ref{subsec:6df_addbiasresults}), giving an expected $\beta$ value of $\beta=0.364$ from Eq. \ref{eq:realspacebeta}. We repeat the same test as performed for the simulated data, comparing the constraints with and without the cross-covariance. We do not expect the tilt of the $f\sigma_8$-$\beta$ contour to change appreciably, but do expect the uncertainty in both parameters to reduce. This is because the gradient of the $f\sigma_8$-$\beta$ contour is the galaxy bias, which we do not expect to significantly affected by the inclusion or exclusion of the cross-covariance.

\begin{figure*}
	\centering
	\begin{subfigure}{0.5\textwidth}
		\includegraphics[width=\columnwidth]{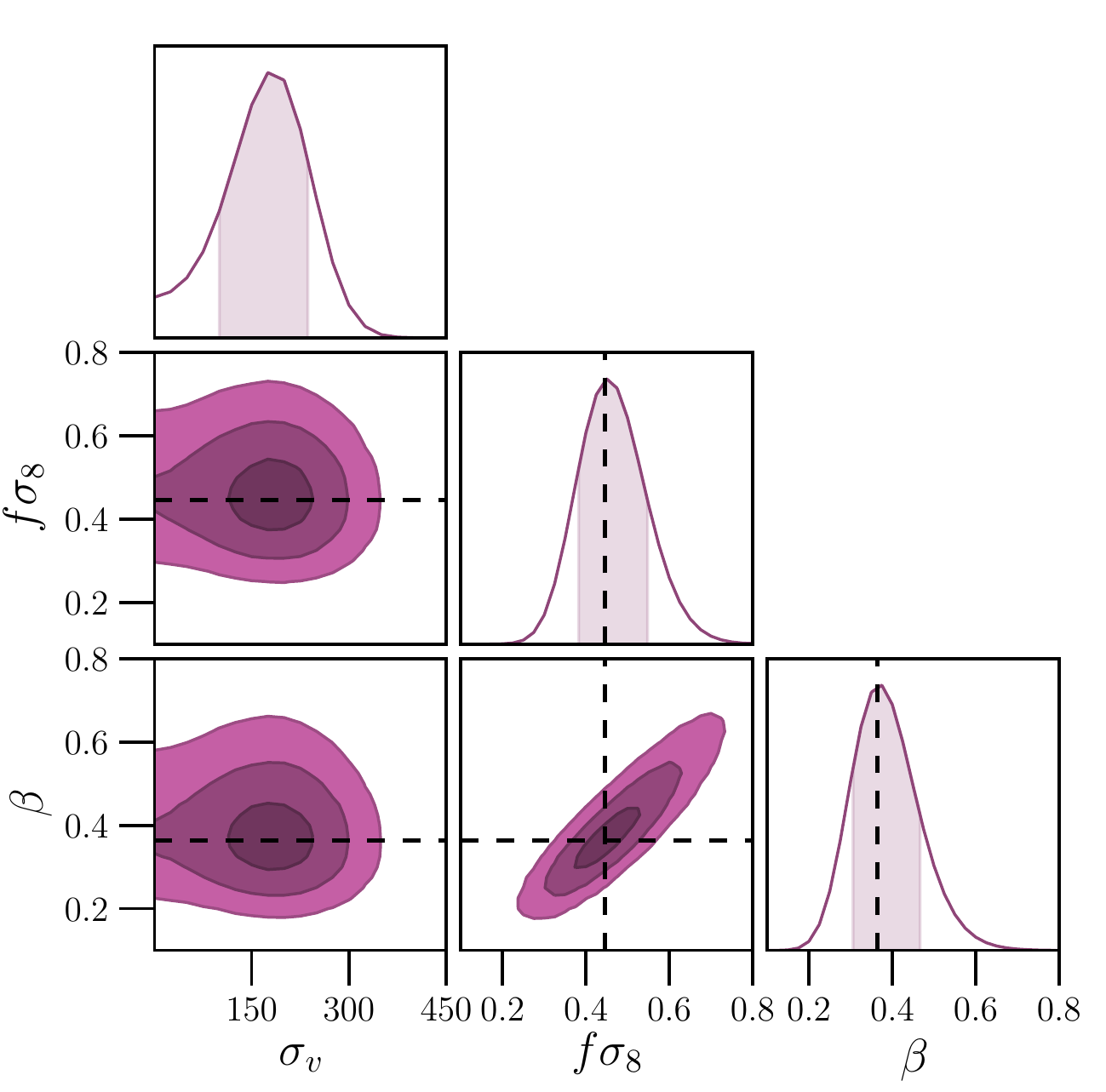}
		\caption{Cross-covariance excluded} \label{fig:6df_sim}
	\end{subfigure}%
	~ 
	\begin{subfigure}{0.5\textwidth}
		\includegraphics[width=\columnwidth]{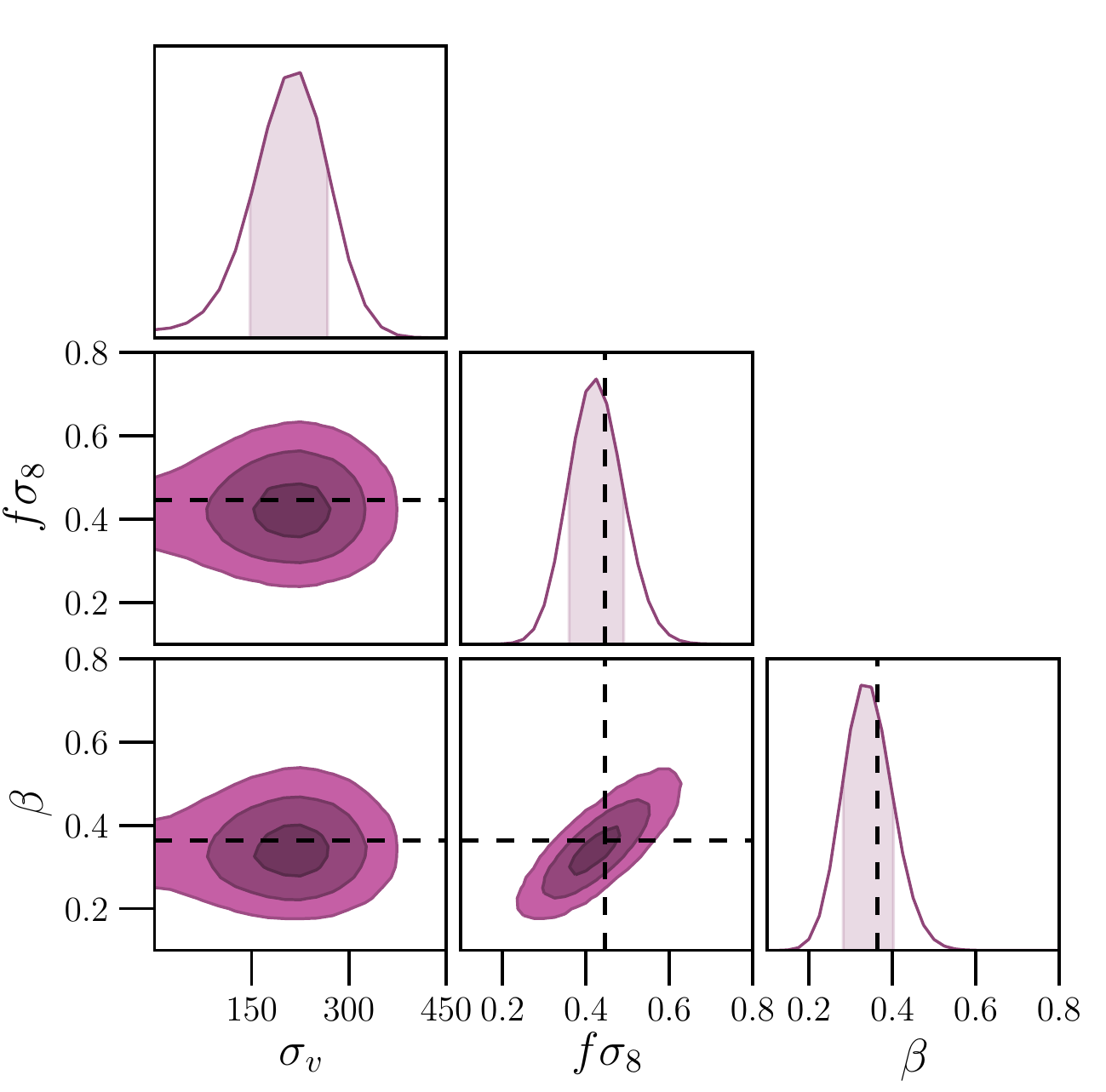}
		\caption{Cross-covariance included}
		\label{fig:6df_joint}
	\end{subfigure}
	\caption{The posteriors of our free parameters, $\sigma_v$ (\kms), $f\sigma_8$ and $\beta$, with shaded 68\% credible intervals, when fitting to the 6dFGS sample. The dashed lines show the expected $f\sigma_8$ and $b_{\rm fit}\sigma_8$ values assuming the fiducial Planck 2015 cosmology (Table \ref{tab:powerspectrummodels}) and an additional bias of $b_{\rm add} = 1.50$.}\label{fig:6df_total}
\end{figure*}

The posteriors for our free parameters when assuming no cross-covariance ($\mathbfss{C}_{g \eta} = \mathbfss{C}_{\eta g} = \mathbfss{0}$) are shown in Fig. \ref{fig:6df_sim}, and the posteriors when including the cross-covariance are shown in Fig. \ref{fig:6df_joint}. The maximum likelihood and median (with 68\% credible intervals) are given in Table \ref{tab:6dfresults}.

\begin{table}
	\centering 
	\caption{Maximum likelihood (ML) and median values (with 68\% credible intervals) of our free parameters for the 6dFGS sample. We include the $\chi^2$ value and $\chi^2$/dof statistic for the maximum likelihood.} \label{tab:6dfresults}
	\begin{tabular}{lcccc} 
		\hline
		&\multicolumn{2}{c}{No cross-covariance}&\multicolumn{2}{c}{Cross-covariance} \\
		& ML & Median & ML & Median \\
		\hline
		$\sigma_v$ (\kms) & 175 & $174^{+63}_{-74}$ &  225 & $210^{+57}_{-63}$ \\
		\\[-0.75em]
		$f\sigma_8$ & 0.425 & $0.461^{+0.087}_{-0.079}$ & 0.400 & $0.424^{+0.067}_{-0.064}$ \\
		\\[-0.75em]
		$\beta$ & 0.350 & $0.380^{+0.087}_{-0.075}$ &0.325& $0.341^{+0.062}_{-0.058}$ \\
		\\[-0.75em]
		$\chi^2$ & 3835.34 & - & 3832.49 & - \\
		$\chi^2/{\rm dof}$ & 0.96 & - & 0.96 & - \\
		\hline        
	\end{tabular}
\end{table}

When including the cross-covariance, we measure $f\sigma_8 = 0.424^{+0.067}_{-0.064}$ which is consistent at the 1$\sigma$ level with the prediction from $\Lambda$CDM using the Planck 2015 cosmological parameters. The fractional uncertainty in $f\sigma_8$ is 18\% when the cross-covariance is ignored and drops to 15\% when the cross-covariance is included. We also measure $\beta = 0.341^{+0.062}_{-0.058}$ which is consistent at the 1$\sigma$ level with our expectation, and $\sigma_v =210^{+57}_{-63}$ \kms \ which is consistent with the literature ($\sigma_v$ usually takes a value between $100$-$300$ \kms). The $\chi^2$/dof value shows that our model is a good fit to the data. We see an improvement in the constraints on both $f\sigma_8$ and $\beta$ when including the cross-covariance, consistent with what we saw from the simulated data. The effect is visually evident in Fig. \ref{fig:6df_total}, where the contour in $\beta$-$f\sigma_8$ space shrinks considerably after adding the cross-covariance. We calculated the Pearson correlation coefficient between $\beta$ and $f\sigma_8$ from our gridded likelihood results, finding $\rho = 0.97$ when excluding the cross-covariance, and $\rho = 0.96$ when including it. Both cases exhibit strong correlation, which is consistent with the fact that $\beta$ is directly proportional to $f\sigma_{8}$ by definition.

When the cross-covariance is introduced, the best-fitting values of $f\sigma_8$ and $\beta$ both decrease, and $\sigma_v$ increases, while the $\chi^2$ statistic slightly decreases. $\sigma_v$ only contributes to the diagonal elements of the peculiar velocity auto-covariance, so any change in this parameter when introducing the cross-covariance will come from its coupling with $f\sigma_8$. If including the cross-covariance lowers $f\sigma_8$, then $\sigma_v$ will increase to compensate for the change in the auto-covariance values along the diagonal.

\subsection{Additional bias} \label{subsec:6df_addbiasresults}
\begin{figure}
	\includegraphics[width=\columnwidth]{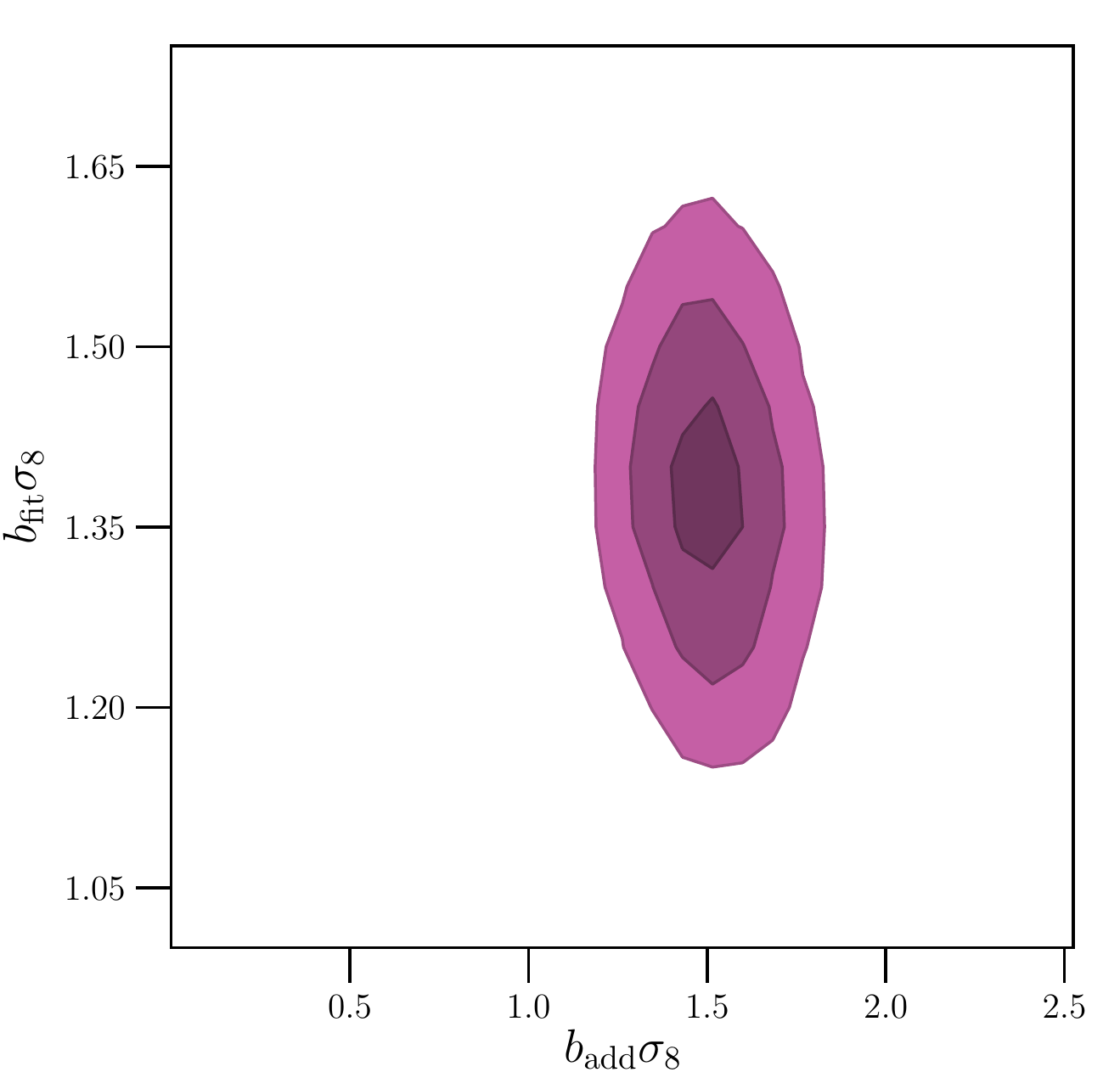}
	\caption{68, 95 and 99\% credible regions of $b_{\rm fit}\sigma_8$ and $b_{\rm add}\sigma_8$ for the 6dFGS galaxy overdensity sample.}
	\label{fig:6df_addbiastest}
\end{figure}

In Section \ref{subsec:integrationbounds} we discussed the additional integral, which increases the value of the galaxy overdensity auto-covariance without directly fitting for the galaxy bias on small scales where non-linear modelling is unreliable. To determine the best-fitting bias for both integrals, we follow the same procedure as for the simulation data (see Section \ref{subsec:gigglezadditionalbias}). The results are given in Fig. \ref{fig:6df_addbiastest}. 

We find median values of $b_{\rm fit}\sigma_8 = 1.38$ and $b_{\rm add}\sigma_8 =1.50$, with a $\chi^2$/dof of 1.10. $b_{\rm add} = 0$ is strongly disfavoured, with $\Delta\chi^2 = 3206.79$. As with the simulation results, we find that the case where $b_{\rm add} = b_{\rm fit}$ is allowed by the data. Again, since we do not wish to include non-linear information, we fix $b_{\rm add}$ at the median value and do not change $k_{\rm max}$.

It is clear from Fig. \ref{fig:6df_addbiastest} that changing $b_{\rm add}$ has little effect on the value of $b_{\rm fit}$. However, since both $b_{\rm fit}$ and $f$ appear in the cross-covariance, it is also important to examine how the value of $b_{\rm add}$ affects our constraints on $f\sigma_8$. We chose three different $b_{\rm add}$ values (1.175, 1.675, 2.375) and ran the full likelihood analysis (using Eq. \ref{eq:likelihoodeq}) for each. The fiducial power spectra for these runs were generated using the Planck 2015 cosmological parameters (see Section \ref{subsec:fidpowerspectrum}).

\begin{figure}
	\includegraphics[width=\columnwidth]{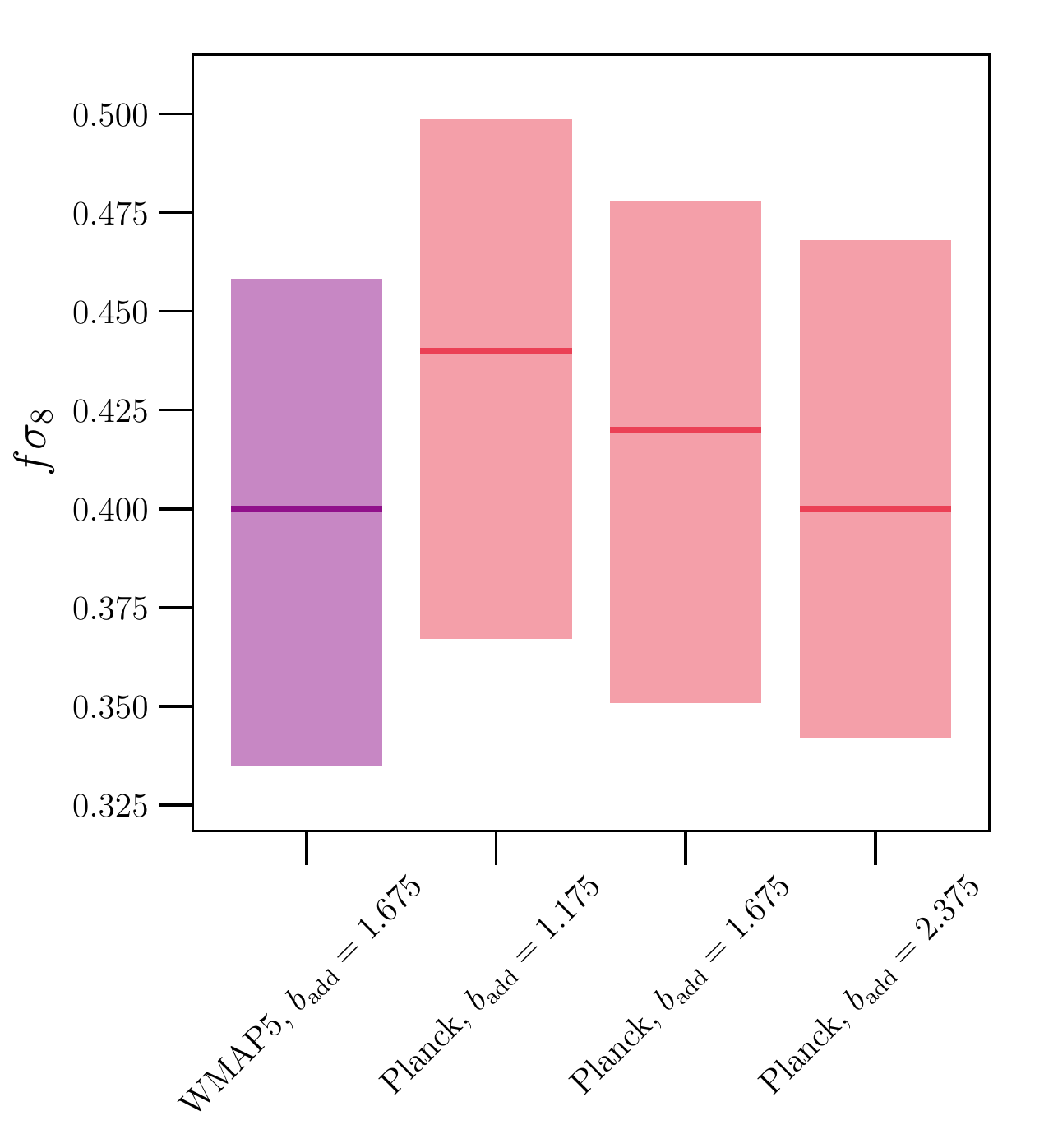}
	\caption{Maximum likelihood (solid bar) and 68\% credible interval (shaded region) of $f\sigma_{8}$ for the 6dFGS sample including the cross-covariance modelling. Results are shown for different values of the additional galaxy bias as well as using both the WMAP five-year and Planck 2015 fiducial cosmologies.}
	\label{fig:6df_changebiastest}
\end{figure}

We also investigated how the choice of cosmological parameters influences our results by adding an additional likelihood run that used the WMAP five-year parameters to generate the fiducial power spectra. We display the 68\% credible intervals around the maximum likelihood for $f\sigma_8$ from the four runs in Fig. \ref{fig:6df_changebiastest}.

When including the additional integral, we find that the value of $b_{\rm add}$ does not significantly influence the constraints on $f\sigma_8$. The maximum likelihood values are all consistent at the 1$\sigma$ level, and the difference between the maximum likelihood value of $f\sigma_{8}$ of the highest and lowest $b_{\rm add}$ is $\sim$0.04, which accounts for around one-third of the statistical error. 

Using the cosmological parameters from WMAP to construct the power spectra gives a lower $f\sigma_{8}$ than using those from Planck for the same additional bias. However, we note that the difference is small, $\sim$0.02 between the maximum likelihood values. Given that the fractional uncertainties in our measurements of $f\sigma_{8}$ are around 15\%, the small contribution of $\sim$3\% from the choice of cosmological parameters is subdominant. However, analyses using data from future surveys will need to be aware that the constraint on $f\sigma_8$ is influenced by the choice of cosmological parameter values used to generate the fiducial power spectra.

\subsection{Direct evidence of cross-covariance} \label{subsec:directevidencecross}
An interesting visualization of our results is to directly plot the analytic cross-covariance for the median values of our free parameters as a function of the separation of grid cells, which can then be compared to an estimate of the cross-covariance present in the 6dFGS data. We represent the estimated covariance as:
\begin{align}
\Sigma_{\delta_g \eta} = \langle \delta_g \eta \rangle - \langle \delta_g \rangle \langle \eta \rangle.
\end{align}
In principle, the average is performed for each pair of cells over many realisations of the data, which would produce a full covariance matrix that could be directly compared to our model. Given that we only have a single realisation of the 6dFGS sample, we instead perform the average for pairs with similar orientation within the survey, specifically their separation, $\bm{r}$, and their angle to the line of sight, $\gamma$ (see Fig. \ref{fig:vectordef}). We split the range $-1 < \cos(\gamma) < 1$ into four equal bins, as this prevents the signals from averaging out. The separation of each pair is assigned to bins of width $20$\mpch\, beginning at $20$\mpch\ and ending at $240$\mpch.

\begin{figure*}
	\includegraphics[width=\textwidth]{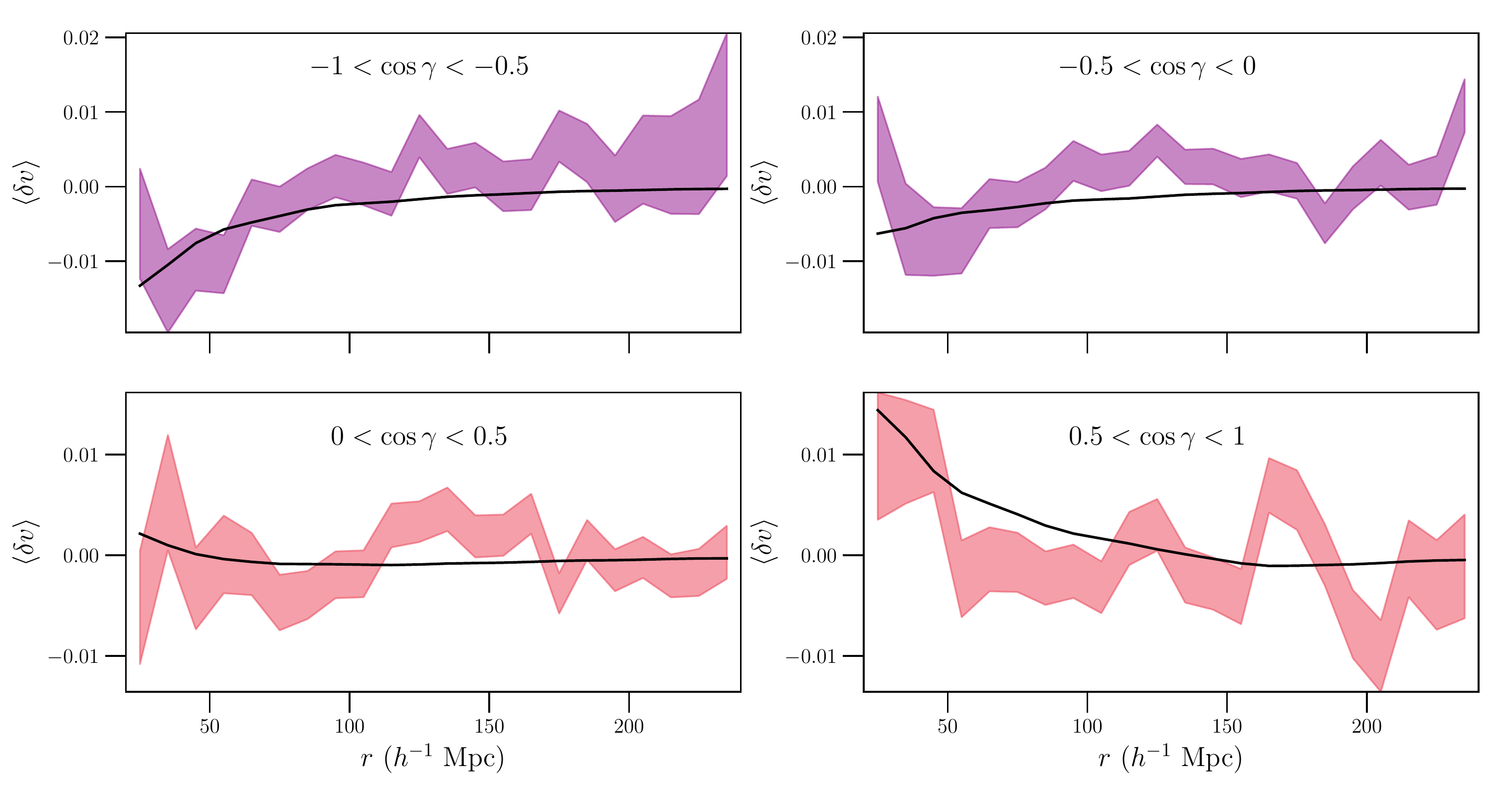}
	\caption{Estimation of the 6dFGS data covariance, $\langle \delta v\rangle$, where the shaded contour represents the covariance plus one standard deviation above and below. The black line shows the prediction from our 6dFGS model covariance for the median values obtained when including the cross-covariance.}
	\label{fig:6df_dv_both}
\end{figure*}

Performing the average for $N$ pairs in each bin produces the covariance as a function of separation for each angular bin, and we estimate the standard deviation of this quantity as:
\begin{align}
\sigma_{\delta_g \eta} = \frac{1}{\sqrt{N}}\sqrt{\langle \delta_g^2 \eta^2 \rangle - \langle \delta_g \eta \rangle^2}.
\end{align}
We average the model covariance matrix for the median values of our free parameters using the same scheme. This can then be directly compared to our estimation from the data, which is shown in Fig. \ref{fig:6df_dv_both}.

We note that the standard deviations will be underestimated since there will be an additional contribution from sample variance that we have not estimated. However, the agreement between our model and the estimated covariance of our data is reassuring. There is good visual evidence from the estimated contours that there is a non-zero cross-correlation between the peculiar velocity field and galaxy overdensity field on separations up to $\sim$50\mpch, and that we are able to successfully model this.

During the production stage for this work, \cite{Nusser2017} presented a direct measurement of the cross-correlation using peculiar velocities from the \textit{cosmicflows-3} catalogue \citep{Tully2016} and galaxy positions from the 2MASS Redshift Survey \cite{Huchra2012}. The also find evidence for a non-zero cross-correlation on scales up to $\sim$50 \mpch.

\subsection{Comparison to previous work}
\subsubsection{Multi-tracer approaches}
Previous theoretical studies have shown that using multiple tracers of the underlying matter overdensity field can improve constraints on cosmological parameters  \citep{Seljak2009,McDonald2009,Bernstein2011,Gil-Marin2010,Abramo2013}. While these works focus on different cosmological tracers and parameters of interest, there are several broad conclusions:
\begin{itemize}
	\item Tracers will contain a statistical error contribution from the sample variance associated with the matter overdensity field 
	\item The sample variance is associated with the volume of space observed and is linked to the number of Fourier modes observed at a given scale
	\item The sample variance contributes to the uncertainty in the power spectrum measurement through $\sigma_P/P = \sqrt{2/N_m}$ for all observable modes, $N_m$, so large scales (small $k$) will be most affected since there are fewer modes for a fixed volume
	\item For two tracers, the fact that they both trace the same matter overdensity field provides two benefits:
	\begin{itemize}
		\item The uncertainty in the ratio of the two tracers does not contain a sample variance contribution and can be precisely known in the limit of no other uncertainty
		\item The cross-power spectrum is available and provides additional constraints. 
	\end{itemize}
\end{itemize}
\cite{Blake2013} presented the first multi-tracer redshift-space distortion (RSD) analysis of observational data, using two galaxy samples with different biases from the Galaxy And Mass Assembly (GAMA) survey. They found a 10-20\% improvement in their measurement of $f\sigma_8$ when utilising the cross-covariance. We see behaviour that is consistent with the expected outcomes from theoretical work and with the results presented by \cite{Blake2013}: our constraints on $f\sigma_8$ and $\beta$ both improve when including the cross-covariance term. The fractional improvement on our measurement of $f\sigma_8$ is 20\%, and the fractional improvement on $\beta$ is 26\%. 

\subsubsection{Forecasts for 6dFGS cross-covariance analysis}
Forecasts of parameter constraints have been performed for multi-tracer analyses that utilise galaxy overdensity and peculiar velocity data. \cite{Koda2014} and \cite{Howlett2017} presented forecasts for 6dFGS that can be compared to our results. 

\cite{Koda2014} performed a Fisher matrix forecast for $f\sigma_8$ and $\beta$ given the properties of the peculiar velocity sample, 6dFGSv. Constraining the parameters out to $k=0.1$\hmpc, they found an expected fractional uncertainty of 25\% on $f\sigma_8$ when only using the peculiar velocity auto-covariance, falling to an expected fractional uncertainty of $15\%$ when including the cross-covariance and galaxy overdensity auto-covariance. The complete covariance also gives an expected fractional uncertainty of 16\% on $\beta$.

\cite{Howlett2017} used the same modelling and formalism as \cite{Koda2014} but considered multiple extensions. For example, they modelled the full 6dFGSz and 6dFGSv samples. They found similar results to \cite{Koda2014}, with the peculiar velocity auto-covariance providing an expected fractional uncertainty of 25.1\% on $f\sigma_8$, and expected fractional uncertainties of 11.2\% on $f\sigma_8$ and 12.3\% on $\beta$ when including the cross-covariance and galaxy overdensity auto-covariance. The tightening of their constraints relative to \cite{Koda2014} can be attributed to the increased number of galaxies from modelling the full 6dFGSz sample.

Our analysis contains some differences in terms of modelling, sample and scales fitted. However, we find that our results are in line with these forecasts. We obtain fractional uncertainties of 15\% on $f\sigma_8$ and 18\% on $\beta$ when using the cross-covariance, and 18\% on $f\sigma_8$ when only using the peculiar velocity auto-covariance. The fractional uncertainties when including the cross-covariance are very similar to those from \cite{Koda2014} although larger than those from \cite{Howlett2017}; this is likely due to their significantly larger galaxy overdensity sample. Our uncertainty from the peculiar velocity auto-covariance is somewhat smaller than forecasted by both studies, in agreement with the results of \cite{Johnson2014} (see below), which we attribute to slight differences in assumptions.

\subsubsection{6dFGS velocity and redshift-space distortion results}
\cite{Johnson2014} constrained $f\sigma_8$ using only the peculiar velocity auto-covariance. When using the 6dFGSv sample, they found $f\sigma_8 = 0.428^{+ 0.079}_{-0.068}$, which has a fractional uncertainty of 17\%. While the fractional uncertainties are similar, our $f\sigma_8$ value is higher when only using the peculiar velocity auto-covariance. However, we find a similar value of $f\sigma_8 = 0.424^{+ 0.067}_{-0.064}$ once cross-covariance is included. This may be due to subtle differences in modelling and approach between our two studies. In particular, \cite{Johnson2014} analytically marginalised over the peculiar velocity zero-point, which would lower their measurement relative to ours. \cite{Huterer2017} also presented an analysis of the peculiar velocity auto-covariance for 6dFGSv. We note that they applied their own Fundamental Plane model to the data. Their constraint of $f\sigma_8 = 0.481^{+0.067}_{-0.064}$ at an effective redshift of z=0.02 is consistent with our result and that from \cite{Johnson2014}.

We can also compare our results to the RSD analysis for 6dFGS, performed by \cite{Beutler2012}, who found $f\sigma_8 = 0.423 \pm 0.055$ at an effective redshift of $z=0.067$, and $\beta = 0.373 \pm 0.054$, with fractional uncertainties of 13\% and 14\% respectively. Both of our constraints are consistent with this work at the 1$\sigma$ level. We do not expect to perfectly recover $\beta$, as our use of a volume-limited sample will preference higher mass halos compared to the sample used by \cite{Beutler2012}, which would increase our galaxy bias relative to theirs. 

Finally, we compare our result to recent measurements of the growth rate from 6dFGS by \cite{Achitouv2017}. They present an RSD analysis of the galaxy-galaxy and galaxy-void correlation functions, utilising realistic mocks to estimate the uncertainties in their results. For the galaxy-galaxy correlation, their result of $f\sigma_8=0.42\pm0.06$ is entirely consistent with that from \cite{Beutler2012}. They also find $f\sigma_8 = 0.39 \pm 0.11$ when analysing the void-galaxy correlation, which has a fractional uncertainty of 28\%. Again, we are consistent at the 1$\sigma$ level with both of these results. See Fig. \ref{fig:constraintscomp} for a visual comparison of the previous 6dFGS $f\sigma_8$ measurements with the measurement from this work. 

\subsubsection{Density-velocity comparison approaches}
The practice of studying the relationship between the observed galaxy overdensity and peculiar velocity fields dates back to the 1990's and has revolved around using gravitational instability to link them (for early examples see \citealt{Kaiser1991} and \citealt{Strauss1992}). The general approach used over the last 20 years involves measuring both fields, applying gravitational instability theory to predict one from the other, and then comparing the prediction and the data to extract constraints on cosmological parameters. The most popular method is a velocity-velocity comparison, where the model peculiar velocity field is constructed from the real-space galaxy overdensity field using Eq. \ref{eq:gravinst}. $\beta = f/b_r$ is then constrained by comparing the model peculiar velocity field to the observed one. For an early review of this topic and different comparison methods, see \cite{Strauss1995}.

We compare our results to $f\sigma_8$ constraints from several representative works: \cite{Pike2005}, \cite{Davis2011} and \cite{Carrick2015}. \cite{Pike2005} and \cite{Carrick2015} used similar methods where they utilised \texttt{VELMOD} (a velocity-velocity comparison approach developed by \citealt{Willick1997}) with some extensions. \cite{Davis2011} expanded both the modelled peculiar velocity field and the observed field in terms of a set of basis functions and then compared the expansion coefficients between the fields. All three analyses used similar data sets, relying on redshifts collected from the 2-Micron All-Sky Survey (with various extensions), and peculiar velocities from the Spiral Field I-Band survey (again with extensions). 

\begin{figure}
	\includegraphics[width=\columnwidth]{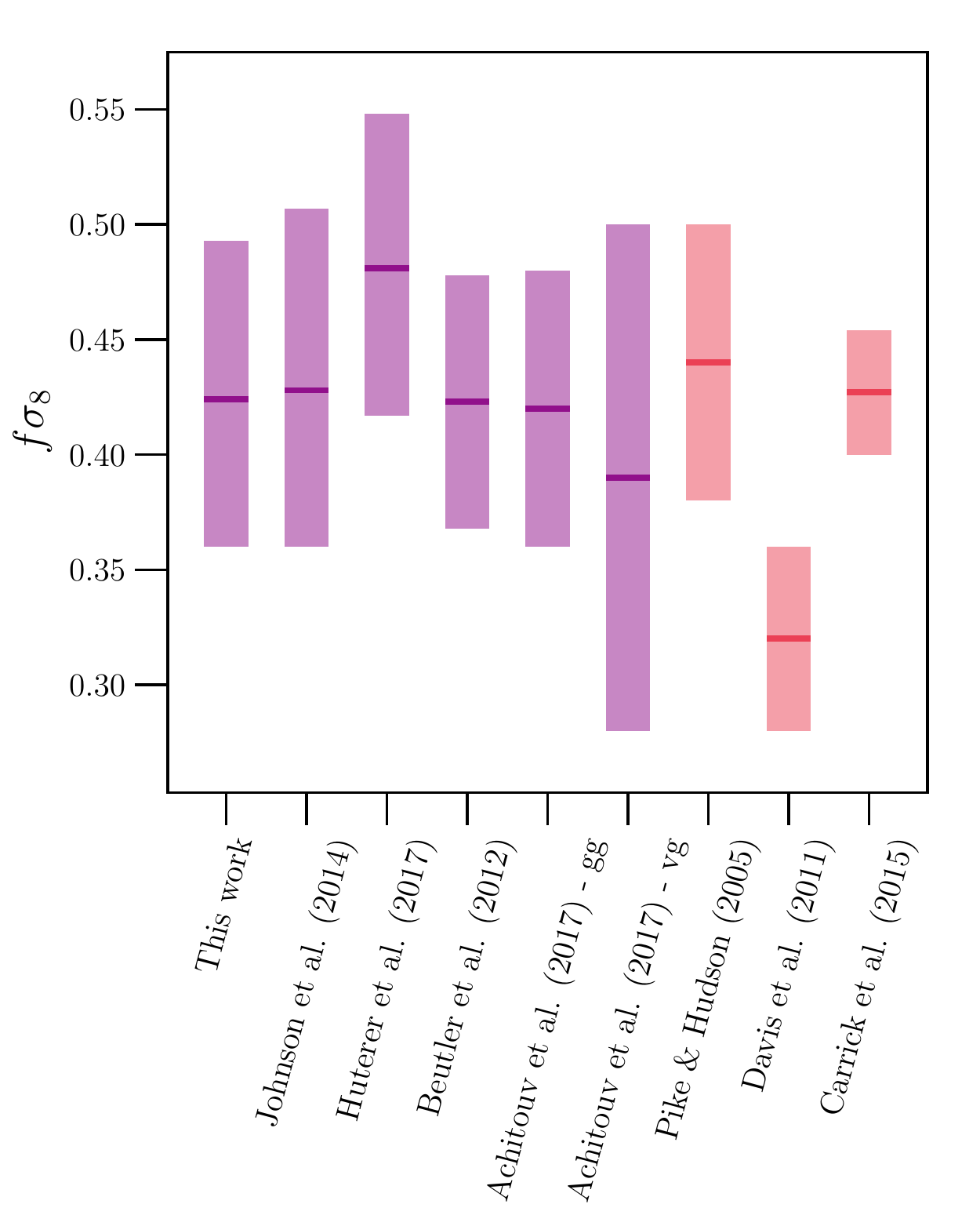}
	\caption{Maximum likelihood (solid bar) and 68\% credible interval (shaded region) of $f\sigma_8$ for this work. Other works from 6dFGS are shown in purple, and velocity-velocity comparisons are shown in pink. For \citet{Achitouv2017}, \textit{gg} corresponds the galaxy-galaxy result, and \textit{vg} corresponds to the void-galaxy result.}
	\label{fig:constraintscomp}
\end{figure}

We do not compare $\beta$ values, as the galaxy overdensity samples will be different between 6dFGS and other surveys, which affects the galaxy bias. We quote the normalised $f\sigma_8$ results, which each survey produces by making an estimate of the galaxy bias for their sample. \cite{Pike2005} found $f\sigma_{8}=0.44 \pm 0.06$, \cite{Davis2011} found $f\sigma_8 = 0.32 \pm 0.04$, and \cite{Carrick2015} found $f\sigma_8 = 0.427 \pm 0.027$. See Fig. \ref{fig:constraintscomp} for a visual comparison of these results with the result from this work.

Our constraint on $f\sigma_8$ when including the cross-covariance is consistent at the 1$\sigma$ level with \cite{Pike2005} and \cite{Carrick2015}, and only just so with \cite{Davis2011}, with has a similar fractional uncertainty to \cite{Pike2005}. The most obvious difference between these approaches and ours is that we model the joint statistics, whereas these approaches make a model from one observation and compare it to the other. They also use more advanced modelling than our current approach: for example, \cite{Carrick2015} implemented a weighting scheme to model luminosity-dependent bias, whereas our linear bias approach is quite simplistic. However, an advantage of our approach is that the modelling can be easily extended by changing the power spectra in the covariance model, and there are multiple improvements that can be explored here. Additionally, the covariance matrix framework makes propagation of errors straightforward, whereas this can be difficult in the comparison approach.

\subsection{Future work} \label{subsec:futurework}
This work has presented the first multi-tracer analysis for galaxy overdensities and peculiar velocities that models the covariance rather than reconstructing the physical fields. This gives it a significant number of advantages, especially its flexibility in testing different cosmological models and its ability for the theory to be easily refined and extended.

While the application of this method to 6dFGS has yielded promising results, more work can be done to better quantify systematics. This can be addressed by working with multiple mocks of 6dFGS, which have a realistic number density for both samples and observational errors. The direct comparison approach presented in Section \ref{subsec:directevidencecross} would also benefit, as more samples will improve our understanding of the sample variance contribution to the 6dFGS result, which cannot be accounted for in the current comparison.

The covariance modelling can also be significantly improved by moving to power spectra that account for RSD, such as those presented by \cite{Koda2014}. This would give a more reliable estimate of $f\sigma_8$ and $\beta$. Additionally, adding RSD serves to further constrain the relationship between $f\sigma_8$ and $\beta$ since the RSD correction is a function of both parameters.

We can also fit for $f\sigma_8$ as a function of scale by splitting the integrals in Eq. \ref{eq:ddcov_gridded_maintext}-\ref{eq:vdcov_gridded_maintext} into multiple bins. Coupling this with the idea that modified gravity models exhibit scale-dependent behaviour in the growth rate of structure provides a new opportunity to test alternative cosmological models. This is especially the case if the modelling is changed to incorporate power spectra from these theories, as using $\Lambda$CDM power spectra only provides a consistency test. 

We are currently working on these improvements and will present them in a follow-up paper. 
	
\section{Summary}\label{sec:summary}
We have presented the first joint statistical analysis of the galaxy overdensity and peculiar velocity fields in which the cosmological physics is modelled in the covariance, extending the work of \cite{Johnson2014}. This includes a complete derivation of the analytic form of the cross-covariance between these two fields and we found its behaviour is consistent with physical intuition. We also found evidence for a non-zero cross-covariance when testing against simulations, as well when we applied our method to data from the 6-degree Field Galaxy Survey. 

For peculiar velocity and galaxy overdensity measurements drawn from 6dFGS, we found the normalised growth rate of structure at redshift zero to be $f\sigma_8 = 0.424^{+0.067}_{-0.064}$, which is consistent with the $\Lambda$CDM prediction for the Planck 2015 cosmological parameters. We also constrained the redshift-space distortion parameter for our sample, finding $\beta=0.341^{+0.062}_{-0.058}$. Our constraint on $f\sigma_8$ improves on that from \cite{Johnson2014} who quoted
a fractional uncertainty of 17\%, where we found a fractional uncertainty of 15\%. This improvement is entirely consistent with the current theory of multi-tracer analyses: including the cross-covariance will improve constraints on the model's cosmological parameters. Our results are also consistent with the redshift-space analyses of 6dFGS by \cite{Beutler2012} and \cite{Achitouv2017}, as well as previous forecasts for 6dFGS from \cite{Koda2014} and \cite{Howlett2017}. Finally, we also saw consistency with alternative methods of analysing the relationship between the galaxy overdensity field and the peculiar velocity field.

This is the first maximum-likelihood fit to the galaxy overdensity and peculiar velocity auto- and cross-covariances, providing a new way of looking at multi-tracer approaches. Given the strong existing theoretical work in this area, it is hugely promising to see concrete evidence of the statistical improvements that are predicted by Fisher matrix forecasting. Importantly, the flexibility offered by modelling the covariance acts as solid insurance that this method will continue to expand and improve, providing increasingly precise and accurate measurements of cosmological parameters as a function of scale and cosmological model. 

Refining this approach is particularly important, as several large peculiar velocity surveys will come online in the next few years. Taipan is forecast to obtain a fractional uncertainty on the growth rate of structure lower than 4\%, and lower still when combined with Wallaby \citep{Howlett2017}. We clearly still have much to gain from large-scale structure data, and taking advantage of the cross-correlation between different tracers is a strong step in improving our knowledge and understanding of the behaviour of our Universe.
	
	
	\section*{Acknowledgements}\label{sec:acknowledgements}
	We are grateful to H\'{e}ctor Gil-Mar\'{i}n for providing an insightful and constructive referee report. The 6dF Galaxy Survey was made possible by contributions from many individuals towards the instrument, the survey and its science. We particularly thank Matthew Colless, Heath Jones, Will Saunders, Fred Watson, Quentin Parker, Mike Read, Lachlan Campbell, Chris Springob, Christina Magoulas, John Lucey, Jeremy Mould, and Tom Jarrett, as well as the dedicated staff of the Australian Astronomical Observatory and other members of the 6dFGS team over the years. We are also grateful to Andrew Johnson and Cullan Howlett for useful discussions. CA thanks Sabine Bellstedt and Geoff Bryan for their input. We have used \texttt{matplotlib} \citep{Hunter2007} for the generation of scientific plots. This research was conducted by the Australian Research Council Centre of Excellence for All-sky Astrophysics (CAASTRO), through project number CE110001020. CA is supported by an Australian Government Research Training Program Scholarship.
	
	
	
	
	\bibliographystyle{mnras}
	\bibliography{cadams_densityvelocitycrosscorrelation6dfgs_sub4} 

	
	
	
	\appendix
	
	\section{Analytic derivation of peculiar velocity and galaxy overdensity auto- and cross-covariances} \label{sec:covariance_derivation}
	In this Appendix we present the derivation of the galaxy overdensity auto-covariance ($\mathbfss{C}_{\delta_g \delta_g}$), peculiar velocity auto-covariance ($\mathbfss{C}_{v_p v_p}$), and the cross-covariances ($\mathbfss{C}_{\delta_g v_p}$ and $\mathbfss{C}_{v_p \delta_g}$). Within this document, we refer to the logarithmic distance ratio parameter $x(z)$ (the notation used in \citealt{Johnson2014}), as $\eta(z)$ (as used in \citealt{Scrimgeour2016}) to avoid confusion with configuration-space positions. We wish to construct these covariances to build the larger covariance:
	\begin{eqnarray}
	\bm{\Delta} =  \begin{pmatrix}
	\bm{\delta} \\ \bm{\eta}
	\end{pmatrix}, \ \ \ \ \mathbfss{C} = \begin{pmatrix}
	\mathbfss{C}_{\delta \delta}  \ \mathbfss{C}_{\delta \eta} \\
	\mathbfss{C}_{\eta \delta} \ \mathbfss{C}_{\eta \eta}
	\end{pmatrix}.
	\end{eqnarray}
	where $\bm{\Delta}$ is the vector containing the list of overdensities, $\bm{\delta}_g$, and logarithmic distance-ratios, $\bm{\eta}$, as measured from the 6-degree Field Galaxy Survey. Here, we use the analytic models in Fourier space for galaxy overdensity, $\delta_g(\bm{k})= b \delta_m(\bm{k})$ and peculiar velocity $\bm{v}(\bm{k}')=-iaHf \theta(\bm{k}')\bm{k}'/k'^2$. The result for peculiar velocity is often quoted without derivation, so we provide a derivation in Appendix \ref{sec:pv_fourier} for reference.
	
	Throughout, we use the following position conventions:
	\begin{align}
	&\bm{x}_s = (x_{s_x}, x_{s_y}, x_{s_z}), \ |\bm{x}_s| = x_s \\
	&\bm{x}_t = (x_{t_x}, x_{t_y}, x_{t_z}), \ |\bm{x}_t| = x_t \\
	&\bm{r} = \bm{x}_t - \bm{x}_s = (r_x, r_y , r_z), \ |\bm{r}|= r \\
	&\hat{\bm{k}} = (\sin \theta \cos \phi, \sin \theta \sin \phi, \cos \theta)
	\end{align}
	
	Within each step, we will use the expansion of the plane wave term into spherical coordinates
	\begin{align}
	e^{i\mathbfit{k}\cdot \mathbfit{r}} = \sum_\ell i^{\ell} (2\ell+1) j_\ell(kr) P_\ell(\hat{\mathbfit{k}}\cdot \hat{\mathbfit{r}} ), \label{eq:planewaveexp}
	\end{align}
	and the orthogonality condition of the spherical harmonic functions
	\begin{align}
	\int \int (-1)^{m'}Y_{\ell,m}Y_{\ell',-m'} \sin\theta d\theta d\phi = \delta_{\ell,\ell'}\delta_{m,-m'} \label{eq:orthogonality}
	\end{align}
	to determine which terms will remain in the expansion. This is the same method as presented by \cite{Ma2011}. We begin with an example of the method by deriving the commonly known result for the galaxy overdensity auto-covariance, before moving on to quote the result for the peculiar velocity auto-covariance, and derive the result for the cross-covariance. 
	
	\subsection{Galaxy overdensity auto-covariance} \label{subsec:d_autocov_derivation}
	Taking the ensemble average of the product of the configuration-space galaxy overdensity field and its complex conjugate:
	\begin{align}
	C_{\delta_g \delta_g} (\bm{x}_s, \bm{x}_t) &= \langle \delta_g(\bm{x}_s) {\delta_g}(\bm{x}_t)^* \rangle \\
	&= b^2 \langle \delta_m(\bm{x}_s)  \overline{\delta_m}(\bm{x}_t) \rangle \\
	&= b^2 \int \frac{1}{(2\pi)^3} e^{-i\bm{k}\cdot \bm{x}_s} \int \frac{1}{(2\pi)^3} e^{i\bm{k}'\cdot \bm{x}_t} \nonumber \\
	&\qquad {} \langle \delta_m(\bm{k}') \delta_m(\bm{k}) \rangle d^3 \bm{k}' d^3 \bm{k}. 
	\end{align}
	Taking $ \langle \delta_m(\bm{k}') \delta_m(\bm{k}) \rangle = (2\pi)^3 P_{mm}(k') \delta_{\rm{D}}^3(\bm{k}'-\bm{k})$ gives:
	\begin{align}
	C_{\delta_g \delta_g} (\bm{x}_s, \bm{x}_t) &= b^2  \int \frac{1}{(2\pi)^3} e^{i\bm{k}\cdot (\bm{x}_t-\bm{x}_s)} P_{mm}(k) d^3 \bm{k} \\
	&= \frac{b^2}{2\pi^2}\int_0^{\infty} P_{mm}(k) k^2 \nonumber \\ 
	&\qquad {} \int_0^{\pi} \int_0^{2\pi} \frac{1}{4\pi} e^{i\bm{k}\cdot \bm{r}}  \sin(\theta) d\phi d\theta dk. \label{eq:ddcov_angular}
	\end{align}
	From Eq. \ref{eq:planewaveexp}, and the fact that there are no other functions within Eq. \ref{eq:ddcov_angular} that can be written in terms of spherical harmonics, the orthogonality condition, Eq. \ref{eq:orthogonality}, ensures that only the $\ell=0$ term of the expansion remains, giving:
	\begin{align}
	C_{\delta_g \delta_g} (\bm{x}_s, \bm{x}_t) &= \frac{b^2}{2\pi^2}\int_0^{\infty} P_{mm}(k) k^2 \nonumber \\
	&\qquad {}  \int_0^{\pi} \int_0^{2\pi} \frac{1}{4\pi} j_0(kr)  \sin(\theta) d\phi d\theta dk \\
	&= \frac{b^2}{2\pi^2}\int P_{mm}(k) k^2 j_0(kr) dk, \label{eq:ddcov_final}
	\end{align}
	which is a familiar result.
	
	\subsection{Peculiar velocity auto-covariance}\label{subsec:v_autocov_derivation}
	\cite{Ma2011} already present the derivation of the peculiar velocity auto-covariance, so here we summarise the result, using the same notation as above for clarity. Unlike the galaxy overdensity auto-covariance in Eq. \ref{eq:ddcov_final}, the peculiar velocity auto-covariance also depends on the angle between the two vectors, defined here as $\cos(\alpha) = \bm{x}_s \cdot \bm{x}_t$.
	
	\begin{align}
	C_{v_p v_p} (\bm{x}_s, \bm{x}_t) &=  \langle v_p(\bm{x}_s) {v_p}(\bm{x}_t)^* \rangle \\
	&= \langle (\bm{v}(\bm{x}_s)\cdot{\hat{\bm{x}}_s}) ({\bm{v}}(\bm{x}_t)^*\cdot{\hat{\bm{x}}_t}) \rangle  \\
	&= \left( aHf \right)^2 \int \frac{1}{(2\pi)^3} \frac{\hat{\bm{k}} \cdot \hat{\bm{x}}_s}{k} e^{-i\bm{k}\cdot\bm{x}_s} \nonumber \\
	&\qquad {}  \int \frac{1}{(2\pi)^3} \frac{(\hat{\bm{k}}'\cdot \hat{\bm{x}}_t)}{k'} e^{i\bm{k}' \cdot \bm{x}_t} \langle  \theta(\bm{k}') \theta(\bm{k}) \rangle d^3 \bm{k}' d^3\bm{k} \\
	&= \left( aHf \right)^2 \int \frac{1}{(2\pi)^3} \frac{(\hat{\bm{k}}\cdot \hat{\bm{x}}_s)(\hat{\bm{k}}\cdot \hat{\bm{x}}_t)}{k^2} \nonumber \\
	&\qquad {} e^{i\bm{k}\cdot (\bm{x}_t-\bm{x}_s)} P_{\theta \theta}(k) d^3 \bm{k} \\
	&= \frac{(aHf)^2}{2\pi^2} \int P_{\theta \theta}(k) \left( \frac{1}{3} \cos \alpha [j_0(kr) - 2j_2(kr) ] \right.  \nonumber \\
	&\qquad {} \left. + \frac{x_s x_t}{r^2} j_2(kr)\sin^2 \alpha \right) dk.
	\end{align}
	
	\subsection{Cross-covariance}\label{subsec:dv_crosscov_derivation}
	We now repeat the process for the cross-covariance for a galaxy overdensity located at $\bm{x}_s$ and peculiar velocity located at $\bm{x}_t$:  
	\begin{align}
	C_{\delta_g v_p} (\bm{x}_s, \bm{x}_t) &= \langle \delta_g(\bm{x}_s) {v_{p}}(\bm{x}_t)^* \rangle  \\
	&= \langle \delta_g(\bm{x}_s) ({\bm{v}}(\bm{x}_t)^{*}\cdot{\bm{\hat{x}}_t}) \rangle  \\
	&= iaHfb \int \frac{1}{(2\pi)^3} e^{-i\bm{k}\cdot \bm{x}_s} \int \frac{1}{(2\pi)^3} \frac{(\hat{k}'\cdot \bm{\hat{x}}_t)}{k'} e^{i\bm{k}' \cdot \bm{x}_t} \nonumber \\
	&\qquad {}  \langle \delta_m(\bm{k}) \theta(\bm{k}')\rangle d^3 \bm{k}' d^3\bm{k} \\
	&= iaHfb \int \frac{1}{(2\pi)^3} \frac{1}{k} P_{m \theta}(k)(\hat{\bm{k}}\cdot \hat{\bm{x}}_t) e^{i\bm{k}\cdot (\bm{x}_t-\bm{x}_s)} d^3 \bm{k} \\
	&= iaHfb \int \frac{1}{2\pi^2} P_{m \theta}(k) k \int \int \frac{1}{4\pi} (\hat{\bm{k}}\cdot \hat{\bm{x}}_t) \nonumber \\ 
	&\qquad {} e^{i\bm{k}\cdot \bm{r}} \sin(\theta) \  d\phi d\theta dk. 
	\end{align}
	To simplify the evaluation of the angular integral, we can express it in terms of spherical harmonics, and utilise the orthogonality condition (Eq. \ref{eq:orthogonality}) as well as the formula for complex conjugation $Y_{l,m}^{*}=(-1)^mY_{l,-m}$. Beginning with the $\hat{\bm{k}}\cdot \hat{\bm{x}}_t$ term:
	\begin{align}
	\hat{\bm{k}}\cdot \hat{\bm{x}}_t &= \frac{1}{x_t} \left( x_{t_x}\sin\theta \cos\phi + x_{t_y}\sin\theta\sin\phi + x_{t_z}\cos\theta \right) \\
	&= \frac{1}{x_t} \left(\sqrt{\frac{2\pi}{3}}x_{t_x}(Y_{1,-1}-Y_{1,1}) \right. \nonumber \\  
	&\qquad {} \left. + i\sqrt{\frac{2\pi}{3}}x_{t_y}(Y_{1,-1}+Y_{1,1}) + 2\sqrt{\frac{\pi}{3}}x_{t_z}Y_{1,0} \right).  \label{eq:kdotx_harmonic}
	\end{align}
	As for the exponential, we can utilise the plane wave expansion in terms of the spherical Bessel functions and Legendre polynomials. Given the form of Eq. \ref{eq:kdotx_harmonic} and the orthogonality property of the spherical harmonic functions, we can see that only the $\ell=0,1$ terms will possibly survive, since $\ell=0$ does not contribute a spherical harmonic term ($P_0(x)=1$), and $\ell \geq 2$ will disappear due to the Kronecker delta $\delta_{\ell,\ell'}$. Thus, we can write the exponential as:
	\begin{align}
	e^{i\bm{k}\cdot \bm{r}} &= j_0(kr) + 3ij_1(kr)P_1(\hat{\bm{k}}\cdot \hat{\bm{r}})  \\
	&= j_0(kr)+ 3ij_1(kr)\frac{1}{r} \left( \sqrt{\frac{2\pi}{3}}r_x(Y_{1,-1}-Y_{1,1}) \right. \nonumber \\
	&\qquad {} \left. + i\sqrt{\frac{2\pi}{3}}r_y(Y_{1,-1}+Y_{1,1})+ 2\sqrt{\frac{\pi}{3}}r_zY_{1,0} \right).
	\end{align}
	According to orthogonality and the conjugation rule, only the $Y_{l,m}Y_{l',m'}$ terms satisfying $m=-m'$ will stay. This reduces the angular equation to:
	\begin{align}
	A &= \int \int \frac{1}{4\pi} (\hat{\bm{k}}\cdot \hat{\bm{x}}_t) e^{i\bm{k}\cdot \bm{r}} \sin(\theta) d\phi d\theta \\
	&= \int \int ij_1(kr) \frac{1}{x_tr} (-x_{t_x}r_{x} Y_{1,-1}Y_{1,1} -x_{t_y}r_{y} Y_{1,-1}Y_{1,1} \nonumber \\
	&\qquad {} +x_{t_z}r_{z} Y_{1,0}Y_{1,0}) \sin(\theta) d\phi d\theta \\
	&=i(\hat{\bm{x}_t}\cdot\hat{\bm{r}})j_1(kr),
	\end{align}
	making the block off-diagonal covariance terms:
	\begin{align}
	C_{\delta_g v_p} (\bm{x}_s, \bm{x}_t) &=\frac{-aHfb}{2\pi^2}\int P_{m\theta}(k) k (\hat{\bm{x}_t}\cdot\hat{\bm{r}})j_1(kr)  dk \\
	C_{v_p \delta_g} (\bm{x}_s, \bm{x}_t) &=\frac{aHfb}{2\pi^2} \int P_{\theta m}(k) k (\hat{\bm{x}_s}\cdot\hat{\bm{r}})j_1(kr)  dk.
	\end{align}
	We note here that the sign of the covariance is conditional on the definition of the vector between the two positions, $\bm{r}$. Since the equation is not symmetric, care must be taken when defining $\bm{r}$. We have found that the easiest way to do this is to define the covariance relative to the galaxy overdensity and peculiar velocity positions, rather than the abstract positions used in the derivation. This means that a single equation can be used for both covariances, without needing to change the sign. For galaxy overdensity at $\bm{x}_{\delta}$ and peculiar velocity at $\bm{x}_v$:
	\begin{align}
	&\bm{r} = \bm{x}_{\delta} - \bm{x}_v \\
	&C_{\delta_g v_p} (\bm{x}_{\delta}, \bm{x}_v) = \frac{aHfb}{2\pi^2} \int P_{m\theta}(k) k (\hat{\bm{x}_v}\cdot\hat{\bm{r}})j_1(kr)  dk \\
	&C_{v_p \delta_g} (\bm{x}_v, \bm{x}_{\delta}) = \frac{aHfb}{2\pi^2} \int P_{\theta m}(k) k (\hat{\bm{x}_v}\cdot\hat{\bm{r}})j_1(kr)  dk.
	\end{align}
	
	\section{Theoretical conversion between peculiar velocity and logarithmic distance ratio} \label{sec:vtoeta_theory}
	The 6-degree Field Galaxy Survey measures peculiar velocities using the Fundamental Plane (see \citealt{Magoulas2012, Springob2014}). The transformation between the measured quantity from the fundamental plane, $\eta=\log_{10}(D_z/D_r)$, to radial peculiar velocity $v_p$ is non-trivial (see \citealt{Springob2014, Scrimgeour2016}). However, the logarithmic distance ratio, $\eta$ has Gaussian uncertainty, and is consequently better suited for the likelihood analysis than peculiar velocity, which has log-normal uncertainty. We opt to update the models presented in Appendix \ref{sec:covariance_derivation} to model $\eta$ instead of radial peculiar velocity.
	
	This has been previously considered for supernovae by \cite{Hui2006}. We note that the convention for 6dFGS peculiar velocities differs from the magnitude variation convention for supernovae, altering the conversion. We now cover the analytic relationship between the radial peculiar velocity, $v_p$, and the observed logarithmic distance ratio, $\eta$. We begin with the definition of $\eta$:
	\begin{align}
	\eta = \log_{10}\left[ \frac{D(z_{obs})}{D(z_H)} \right], \label{eq:eta_definition}
	\end{align}
	where $D$ is the comoving distance in \mpch\, calculated at the observed redshift, $z_{\rm obs}$, and expansion redshift, $z_{H}$. If the Hubble constant is known as a function of redshift (generally assumed as a part of the model), we can express the comoving distance as:
	\begin{align}
	D(z) = c \int_0^z \frac{dz'}{H(z')}
	\end{align}
	Since we do not know $z_H$ we cannot directly evaluate Eq. \ref{eq:eta_definition}. We can however perform a Taylor expansion of $D(z_{\rm obs})$ around $z_H$:
	\begin{align}
	D(z_H) = D(z_{\rm obs}) + \frac{c}{H(z_{\rm obs})}(z_H - z_{\rm obs}).
	\end{align}
	Then, using the relationship between the observed, expansion, and peculiar velocity redshifts: $(1+z_{\rm obs}) = (1+z_H)(1+z_{v_p})$ (where $z_{v_p} = v_p/c$), we can express the ratio between the comoving distances as:
	\begin{align}
	\frac{D(z_H)}{D(z_{\rm obs})} &= 1 + \frac{c}{D(z_{\rm obs})H(z_{\rm obs})}(z_H - z_{\rm obs}) \\
	&= 1 + \frac{c(1+z_{\rm obs})}{D(z_{\rm obs})H(z_{\rm obs})} \left [\frac{1}{(1+v/c)} - 1 \right].
	\end{align}
	Applying a Taylor series to the bracketed term around $v/c = 0$,
	\begin{align}
	\frac{D(z_H)}{D(z_{\rm obs})} &= 1 - \frac{(1+z_{\rm obs})}{D(z_{\rm obs})H(z_{\rm obs})}v. \label{eq:macluarin_ratio}
	\end{align}
	We can then calculate $\eta$ as a function of the radial peculiar velocity by combining Eq. \ref{eq:eta_definition} and Eq. \ref{eq:macluarin_ratio}:
	\begin{align}
	\eta &= -\log_{10}\left[ \frac{D(z_H)}{D(z_{\rm obs})} \right] \\
	&= -\log_{10}\left[1 - \frac{(1+z_{\rm obs})}{D(z_{\rm obs})H(z_{\rm obs})}v\right] \\
	&=-\frac{1}{\ln(10)} \ln\left[ \frac{D(z_H)}{D(z_{\rm obs})} \right] \\
	&=-\frac{1}{\ln(10)} \ln \left[1 - \frac{(1+z_{\rm obs})}{D(z_{\rm obs})H(z_{\rm obs})}v \right].
	\end{align}
	This can then be simplified further from the first order of the Maclaurin series  $\ln(1-x) \approx -x$:
	\begin{align}
	\eta &= \frac{1}{\ln(10)} \frac{(1+z_{\rm obs})}{D(z_{\rm obs})H(z_{\rm obs})}v.
	\end{align}
	We note that this final approximation will fail at very low redshift.
	
	\section{Peculiar velocity field in Fourier space} \label{sec:pv_fourier}
	\subsection{Velocity and matter overdensity in configuration space} \label{subsec:dvconfigspace}
	The velocity field is related to the matter overdensity field through the continuity equation:
	\begin{align}
	\frac{\partial}{\partial t} \rho(\bm{x},t) = -\nabla \cdot \bm{v}(\bm{x},t)\rho(\bm{x},t),
	\end{align}
	where $\rho$ is the average background matter density. Applying first order perturbation theory, we find:
	\begin{align}
	\frac{\partial}{\partial t}\delta_m(\bm{x},t) = -\frac{1}{a}\nabla \cdot \bm{v}(\bm{x},t),
	\end{align}
	where $\delta_m(\bm{x},t)$ is the matter overdensity field and $a$ is the scale factor. Since we are working with a partial differential equation, we may rewrite the matter overdensity field as:
	\begin{align}
	\delta_m(\bm{x},t) = \delta_m(\bm{x})D_1(t) + \delta_m(\bm{x})D_2(t),
	\end{align}
	where $D_1(t)$ describes growth, and $D_2(t)$ describes decay. At late times, the growth mode dominates, leaving:
	\begin{align}
	\delta_m(\bm{x})\frac{d}{d t}D_1(t) = -\frac{1}{a}\nabla \cdot \bm{v}(\bm{x},t). \label{eq:deltavrelation_growthmode}
	\end{align}
	Applying the chain rule, we may express the derivative as:
	\begin{align}
	\frac{d}{d t} &= \frac{d a}{d t} \frac{d \ln(a)}{d a} \frac{d}{d \ln(a)} \\
	&= \dot{a} \frac{1}{a} \frac{d}{d \ln(a)} \\
	&= H\frac{d}{d \ln(a)},
	\end{align}
	where $H$ is the Hubble constant. We may now express Eq. \ref{eq:deltavrelation_growthmode} as:
	\begin{align}
	\nabla \cdot \bm{v}(\bm{x},a) &= - aH\delta_m(\bm{x})\frac{d}{d \ln(a)}D_1(a).
	\end{align}
	It is common to express this final derivative in terms of the growth rate of structure, $f$, where:
	\begin{align}
	f \equiv \frac{d \ln(D_1(a))}{d \ln(a)} = \frac{1}{D_1(a)} \frac{d}{d \ln(a)}D_1(a).
	\end{align}
	Substituting this we find:
	\begin{align}
	\nabla \cdot \bm{v}(\bm{x},a)&= - aHf\delta_m(\bm{x})D_1(a) \\
	&= - aHf\delta_m(\bm{x},a). \label{eq:vdconfigspace}
	\end{align}
	We will be working in the low-redshift universe, where time evolution is negligible. For this reason, we ignore the scale factor dependence in both the velocity and matter overdensity fields.
	
	\subsection{Expressing the velocity in terms of a scalar field in Fourier space}
	The velocity field is a vector field, and the Helmholtz decomposition theorem states that a vector field can be written as the sum of a gradient of a scalar field, and the curl of a vector field. That is:
	\begin{align}
	\bm{v}(\bm{x}) = -\nabla\phi(\bm{x}) + \nabla\times\bm{w}(\bm{x}).
	\end{align}
	Assuming the field is irrotational, we can remove the curl term. We note that this is a long-standing assumption, see \cite{Strauss1995}, Section 7.5. This leaves:
	\begin{align}
	\bm{v}(\bm{x}) = -\nabla\phi(\bm{x}). \label{eq:velconfigspace}
	\end{align}
	We may now evaluate the velocity field in Fourier space as the Fourier transform of Eq. \ref{eq:velconfigspace}:
	\begin{align}
	\bm{v}(\bm{k}) = - \iiint [\nabla \phi(\bm{x})] e^{i\bm{k}\cdot\bm{x}} d^3\bm{x}. \label{eq:velfourierspace}
	\end{align}
	Letting $g(\bm{k},\bm{x}) = e^{i\bm{k}\cdot\bm{x}}$, we can rewrite the integrand using the chain rule:
	\begin{align}
	\nabla(g\phi) &= (\nabla g)\phi + g(\nabla \phi)\\
	-g(\nabla \phi) &= (\nabla g)\phi - \nabla(g\phi). \label{eq:velchainrule}
	\end{align}
	Substituting Eq. \ref{eq:velchainrule} into Eq \ref{eq:velfourierspace} we find:
	\begin{align}
	\bm{v}(\bm{k}) &= \iiint \nabla \left(e^{i\bm{k}\cdot\bm{x}} \right) \phi(\bm{x}) d^3\bm{x} \nonumber \\
	&\qquad \qquad {} {} {} - \iiint \nabla \left(e^{i\bm{k}\cdot\bm{x}} \phi(\bm{x}) \right) d^3\bm{x} \label{eq:velint_twoterms}
	\end{align}
	We note that a variant of Gauss's theorem can be used to rewrite the second term. For a scalar field, $f(\bm{x})$:
	\begin{align}
	\iiint_V \nabla f(\bm{x}) d^3\bm{x} = \iint_S f(\bm{x}) \hat{\bm{n}} dS,
	\end{align}
	where $\bm{n}$ is the vector normal to the surface. Eq. \ref{eq:velint_twoterms} then becomes:
	\begin{align}
	\bm{v}(\bm{k}) &= \iiint \nabla \left(e^{i\bm{k}\cdot\bm{x}} \right) \phi(\bm{x}) d^3\bm{x} - \iint_S e^{i\bm{k}\cdot\bm{x}} \phi(\bm{x}) dS
	\end{align}
	Integrals in this form converge to zero as long as the integrand, $f(\bm{x})$, falls off faster than $1/x$ as $x$ goes to infinity. We take this to be true for $e^{i\bm{k}\cdot\bm{x}} \phi(\bm{x})$, leaving:
	\begin{align}
	\bm{v}(\bm{k}) &= \iiint \nabla \left(e^{i\bm{k}\cdot\bm{x}} \right) \phi(\bm{x}) d^3\bm{x}.
	\end{align}
	The gradient of the exponential function evaluates to:
	\begin{align}
	\nabla e^{i\bm{k}\cdot \bm{x}} = i\bm{k} e^{i\bm{k}\cdot\bm{x}},
	\end{align}
	leaving:
	\begin{align}
	\bm{v}(\bm{k}) &= i\bm{k}\iiint \phi(\bm{x}) e^{i\bm{k}\cdot\bm{x}} d^3\bm{x} \\
	&= i\bm{k} \phi(\bm{k}).
	\end{align}
	
	\subsection{Velocity and matter overdensity in Fourier space} \label{subsec:dvfourierspace}
	We may now use the results from the previous two sections to express the velocity field in terms of the matter overdensity field in Fourier space. Combining Eq. \ref{eq:vdconfigspace} and \ref{eq:velconfigspace}:
	\begin{align}
	&\nabla \cdot (-\nabla \phi(\bm{x})) = -aHf\delta_m(\bm{x}). \\
	\end{align}
	We can then express each field in terms of its Fourier transform:
	\begin{align}
	& \frac{-\nabla^2}{(2\pi)^3}\iiint \phi(\bm{k}) e^{-i\bm{k}\cdot\bm{x}} d^3\bm{k} = \frac{-aHf}{(2\pi)^3}\iiint \delta_m(\bm{k}) e^{-i\bm{k}\cdot\bm{x}} d^3\bm{k}.
	\end{align}
	The gradient operator then only applies to the exponential, which gives:
	\begin{align}
	& \iiint k^2 \phi(\bm{k}) e^{-i\bm{k}\cdot\bm{x}} d^3\bm{k} = -aHf\iiint \delta_m(\bm{k}) e^{-i\bm{k}\cdot\bm{x}} d^3\bm{k}.
	\end{align}
	Which leads to:
	\begin{align}
	k^2 \phi(\bm{k}) &= -aHf\delta_m(\bm{k}) \\
	 \bm{v}(\bm{k}) &= \frac{-i\bm{k}aHf}{k^2}\delta_m(\bm{k}).
	\end{align}



	\bsp    
	\label{lastpage}
\end{document}